\documentclass[twocolumn,english,aps,pra,superscriptaddress,showpacs]{revtex4}
\usepackage{times}
\usepackage[T1]{fontenc}
\usepackage[latin1]{inputenc}
\usepackage{color}
\usepackage{graphicx}
\usepackage{amssymb}

\makeatletter

\usepackage{color}
\usepackage{graphics}

\renewcommand{\vec}{\mathbf}

\usepackage{babel}
\makeatother
\begin{document}

\title{Spontaneous recoil effects of optical pumping on trapped atoms}

\author{S. Wallentowitz}

\affiliation{Facultad de Física, Pontificia Universidad Católica de Chile, Casilla
306, Santiago 22, Chile}

\author{P.E. Toschek}

\affiliation{Institut für Laser-Physik, Universität Hamburg, Jungius-Straße 9,
D-20355 Hamburg, Germany}

\begin{abstract}
The recoil effects of spontaneous photon emissions during optical
pumping of a trapped three-level atom are exactly calculated. Without
resort to the Lamb-Dicke approximation, and considering arbitrary
detuning and saturation of the pump laser, the density of recoil shifts
in phase space is derived. It is shown that this density is not of
Gaussian shape, and that it becomes isotropic in phase space only
for a branching ratio corresponding to fluorescence scattering but
unfavorable for optical pumping. The dependence of its anisotropy
on the laser saturation is discussed in the resonant case, and the
mapping of moments of the atom's center-of-mass motion due to the
pumping is presented. Moreover, it is shown how optimum parameters
for protecting the center-of-mass quantum state from pump-induced
disturbance depend on the specific property to be protected. 
\end{abstract}

\date{August 8, 2008}


\maketitle

\section{Introduction}

Optical pumping (OP) \cite{kastler,kastler2} has proved to be a very
powerful method for preparing specific electronic states of atoms.
By application of a near resonant light field the atoms are excited
and subsequently decay into the desired electronic state via spontaneous
emission. In bulk media such as a solid, or a gas of atoms or molecules,
optical pumping can be understood solely in terms of electronic or
optical properties of the medium. Although in dense gases modifications
arise due to atom-atom interactions mediated by the radiation field,
the mechanical effects of optical pumping may be safely neglected
in such situations, since momentum is transferred to the collective
sample. However, any light interaction with free or weakly bound atoms
not only is accompanied by a change of the internal electronic quantum
state of the atoms, but unavoidably also affects their center-of-mass
motion \cite{kapitza-dirac,ashkin}. Examples of this type of interaction
are found with the momentum diffusion in fluorescence scattering of
free atoms \textcolor{red}{}\cite{frisch,pusep,berman,mandel,letokhov-review},
with the laser cooling of trapped atoms \cite{haensch,letokhov,andreyev,nobelprize,metcalf}
or ions \cite{neuhauser,wineland}. The mechanical effects are particularly
apparent when we consider the interaction of a laser field with an
individual particle, e.g., with a single trapped ion \cite{toschek-sion},
with single atoms confined in a micro cavity \cite{boozer,nussman},
as well as in atom optics \cite{meystre} and laser-control of atoms
\cite{letokhov-book}. When OP is applied to such a system, one has
explicitly to take into account the mechanical effects of the laser-atom
interaction. Sometimes, as with OP of low-lying vibronic states of
weakly bound trapped ions, the radiative recoil generates a desired
effect: cooling of the sample. However, more often than not, the mechanical
effect modifies the momentum or energy distribution of the sample
and the concomitant radiative spectrum in an undesirable way. In particular,
the incoherent nature of spontaneous emission makes it add decoherence
\cite{joos} to the quantum state of the center-of-mass motion of
the atoms. In view of the possible application of such systems to
quantum-state engineering \cite{blatt,vaser}, quantum logic, and
computing \cite{beth,steane,sasura}, the introduction of any such
decoherence is obviously problematic. Thus, the identification and
avoidance of decoherent effects is of utmost importance.

From the experimental viewpoint, the optimum parameters must be found
in order to obtain smallest decoherence. Given a certain atomic species
confined in a trapping potential, several parameters are predetermined
such as the electronic transition frequencies, the relaxation rates,
and the trap frequency. Apart from modifying the trap, the adjustable
parameters are the laser intensity and detuning. One may consider
convenient to use fast optical pumping, faster than the vibrational
period, since the negligible center-of-mass motion improves the efficiency
of the pump process. On the other hand, the required laser intensity
may saturate and broaden the employed atomic transition, which spoils
OP's resolution of vibronic states. Thus, the interplay of saturation
and pumping time requires close inspection. 

The recoil upon atoms interacting with light has been treated in various
contexts, such as for example in the light interaction of free \cite{letokhov-review}
and trapped single atoms \cite{meschede} or in the collective atomic
recoil laser \cite{bonifacio,courteille}. The applied models consider
an atomic two-level system continuously irradiated by near-resonant
monochromatic light. Such a treatment of free atoms leads to a Gaussian
momentum distribution in accordance with the central limit theorem
applied to the statistically independent individual photon recoils
\cite{mandel}. For trapped atoms this situation is different since
here the waiting-time distribution of subsequent spontaneous photon
emission correlates the individual photon recoils in phase space,
and the validity of the central-limit theorem is doubtful. Additionally,
OP requires a third atomic level, a lower target state. The process
is intrinsically terminated by the spontaneous decay into this state.
Thus, contributions with a small number of radiative events may dominate
the interaction. The recoil effects of these contributions make the
resulting distribution of the atom in vibrational phase space substantially
deviate from being Gaussian - the more so, the less is the number
of interaction events with the pump light. 

In this paper, we address a single atom trapped in a harmonic potential.
We model the motion of the atom's center-of-mass by a 1D quantum-mechanical
oscillation along a certain direction. This model can be viewed as
a prototype system that provides us with the essential information
to be applied to one of the real systems mentioned above. Using a
quantum trajectory approach we obtain analytic results for the quantum
statistical properties of the atomic center-of-mass motion. Moreover,
we obtain insight into the decoherence properties by employing a phase-spatial
picture, that lends to illustrative interpretation.

The method of quantum trajectories \cite{qt-hegerfeldt,qt-gardiner,qt-dalibard,qt-carmichael,qt-molmer,qt-garraway,qt-plenio}
is perfectly matched to studying OP with proper inclusion of recoil
effects, and avoiding the Lamb-Dicke approximation: Firstly, the trail
of spontaneously emitted photons of a single experimental realization
is finite, since a stationary state is reached when the system is
no longer affected by the laser drive. This fact simplifies the structure
of single quantum trajectories. Secondly, when using waiting-time
distributions, easy connection is made between the center-of-mass
vibration and the statistics of photon emissions. Finally, an elegant
formulation of recoil effects is found that characterizes the process
of optical pumping, from the viewpoint of the vibrational motion,
by just a single distribution function.

In Sec. \ref{sec:2} we introduce the model system under consideration
and develop its evolution as a quantum trajectory in phase space.
Section \ref{sec:3} outlines the vibronic state of the pumped atom
in terms of the density of recoil shifts in phase space. This density
is characterized in Sec. \ref{sec:4} by its statistical moments that
are expressed by quantities derived from photon-counting statistics.
Section \ref{sec:5} is then devoted to a discussion of the results
in limiting cases, such as fluorescence scattering and maximally anisotropic
scattering in phase space, as well as to optimum laser parameters
in order to minimize detrimental recoil effects of OP. Section \ref{sec:Motivation-and-Context}
places the results in the context of possible observations. Finally,
Sec. \ref{sec:6} provides a summary and conclusions.

\section{Quantum trajectories of optical pumping \label{sec:2}}

\subsection{Quantum master equation \label{sec:2.1}}

\begin{figure}
\begin{centering}\includegraphics[width=0.45\textwidth]{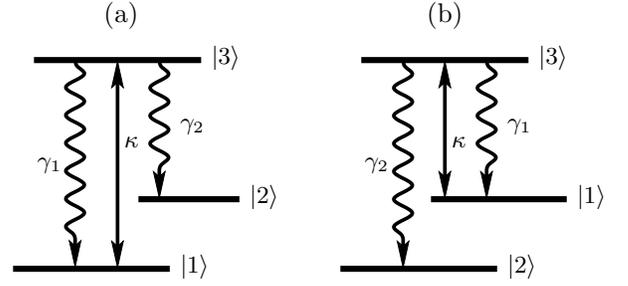}\par\end{centering}

\caption{Draft excitation schemes for the inversion of the two-level system
$|1\rangle\rightarrow|2\rangle$ by optical pumping via the third
auxiliary level $|3\rangle$. Scheme (b) is identical to (a) with
the ground state being replaced by a metastable state.}

\label{fig:scheme}
\end{figure}

Our generic model for treating the motional effects of OP consists
of a $\Lambda$-type three-level atom whose center-of-mass coordinate
is bound in a harmonic trap potential. In this way, photonic-recoil
effects can be described for various systems, such as single ions
in rf traps or neutral atoms in optical-dipole or magneto-optical
traps. The level scheme and setup of optical pumping is shown in Fig.
\ref{fig:scheme}. 

In both equivalent schemes, a laser field is assumed to drive the
electronic dipole transition $|1\rangle\leftrightarrow|3\rangle$,
whereupon photons are spontaneously emitted connected with electronic
dipole relaxation from the auxiliary level $|3\rangle$ to both $|1\rangle$
and $|2\rangle$. The latter process occurs only once, since in state
$|2\rangle$ the atom decouples from the driving laser and thus has
reached is final electronic state. Both level schemes (a) and (b)
include pumping of population from the electronic state $|1\rangle$
to state $|2\rangle$, with the corresponding direct transition being
dipole forbidden. The optical-pumping rate is determined by the electronic
relaxation rates $\gamma_{1}$ and $\gamma_{2}$ of the two dipole
transitions, cf. Fig. \ref{fig:scheme}, and by the laser Rabi frequency
$\kappa=-2dE/\hbar$. Here, $d$ is the electronic dipole moment of
the transition $|1\rangle\leftrightarrow|3\rangle$, and $E$ is the
electric-field amplitude of the laser.

The subject to be addressed is the atom's center-of-mass vibration
at frequency $\nu$ along one principal axis of the trapping potential.
Furthermore, only the motional recoil effects of spontaneous photon
emissions are of interest here. The additional recoil effects due
to the laser interaction have been discussed elsewhere \cite{laser-recoil}
and can be separated by choosing the propagation direction of the
laser beam perpendicular to the direction of motion under consideration.
This motion may be thought to extend along the axis of a linear trap,
while the orthogonal confinement of the particle be \char`\"{}strong\char`\"{},
such that the trap absorbs the momentum transmitted by the laser. 

The dynamics of the vibronic density operator $\hat{\varrho}(t)$
is given by the quantum master equation \cite{laser-recoil} \begin{eqnarray}
\frac{\partial\hat{\varrho}}{\partial t} & = & \frac{1}{i\hbar}[\hat{H}_{0}+\hat{V}(t),\hat{\varrho}]\label{eq:mst1}\\
 &  & +\sum_{a=1,2}\gamma_{a}\left[\hat{\sigma}_{a,-}\left(\hat{\mathcal{R}}_{a}\hat{\varrho}\right)\hat{\sigma}_{a,+}-\frac{1}{2}\left\{ \hat{\sigma}_{a,+}\hat{\sigma}_{a,-},\hat{\varrho}\right\} \right],\nonumber \end{eqnarray}
\textcolor{black}{where $\{\hat{A},\hat{B}\}=\hat{A}\hat{B}+\hat{B}\hat{A}$
denotes the anti-commutator, and $\hat{H}_{0}$ describes} the free
time evolution of the atom in the trap potential %
\footnote{\textcolor{black}{Without loss of generality, the zero-energy level
has been shifted here by $-\hbar(\omega_{1}+\omega_{2})$ for simplicity.}%
}, \begin{equation}
\hat{H}_{0}=\hbar\nu\hat{b}^{\dagger}\hat{b}+\sum_{a=1,2}2\hbar\omega_{a}\hat{\sigma}_{a,z},\label{eq:H0}\end{equation}
with $\omega_{a}$ ($a=1,2$) being the electronic transition frequencies.
Moreover, $\hat{b}$ and $\hat{b}^{\dagger}$ are the annihilation
and creation operators for the vibration along the chosen principal
axis of the trap, and the electronic spin-type operators are given
by\begin{eqnarray}
\hat{\sigma}_{a,-} & = & |a\rangle\langle3|,\qquad\hat{\sigma}_{a,+}=\hat{\sigma}_{a,-}^{\dagger},\label{eq:sigma-a-mp}\\
\sigma_{a,z} & = & \frac{1}{2}\left(|3\rangle\langle3|-|a\rangle\langle a|\right).\label{eq:sigma-a-z}\end{eqnarray}
Finally, the external drive of the laser with frequency $\omega$
is described by the Hamiltonian \begin{equation}
\hat{V}(t)=\frac{1}{2}{\textstyle \hbar\kappa\hat{\sigma}_{1,+}e^{-i\omega t}+{\rm h.c.},}\label{eq:Hl}\end{equation}
with $\kappa=2\vec{d}_{31}\cdot\vec{E}/\hbar$ being the Rabi frequency,
$\vec{d}_{31}=\langle3|\hat{\vec{d}}|1\rangle$ being the transition
dipole moment, and $\vec{E}$ being the electric-field amplitude of
the laser. 

In Eq. (\ref{eq:mst1}) the super-operators $\hat{\mathcal{R}}_{a}$
are responsible for the momentum recoil due to spontaneous emission
of a photon via the electronic decay channels $|3\rangle\rightarrow|a\rangle$
($a=1,2$). They act on an arbitrary operator $\hat{A}$ as \begin{equation}
\hat{\mathcal{R}}_{a}\hat{A}=\int_{-1}^{1}ds\mu_{a}(s)e^{isk_{a}\hat{x}}\hat{A}e^{-isk_{a}\hat{x}},\qquad(a=1,2)\label{eq:recoil-op}\end{equation}
with the respective photon-momentum recoil $\hbar k_{a}=\hbar\omega_{a}/c$
($a=1,2$) being projected on the considered axis according to the
normalized dipole radiation characteristics,\begin{equation}
\mu_{a}(s)=\frac{3}{8\pi}\int_{0}^{2\pi}d\phi\left\{ 1-[\vec{n}_{a}\cdot\vec{n}(\Omega)]^{2}\right\} ,\label{eq:dipole-charac}\end{equation}
where $s=\cos\theta$ is the projection of the photon-emission direction
on the considered motional axis. The unit-normalized vectors $\vec{n}_{a}$
and $\vec{n}(\Omega)$ point into the directions of the transition
dipole moment $\langle3|\hat{\vec{d}}|a\rangle$ and of the spontaneous
emission $\Omega=(\theta,\phi)$, respectively. Assuming, for example,
transition dipole moments orthogonal to the direction of motion under
consideration, Eq. (\ref{eq:dipole-charac}) reduces to\begin{equation}
\mu_{a}(s)=\frac{3}{8}\left(1+s^{2}\right).\label{eq:dipole-charac-perp}\end{equation}

The strength of the recoil effect of individual spontaneous photon
emissions can be characterized by Lamb--Dicke parameters: The position
operator $\hat{x}$ is expressed in terms of the annihilation and
creation operators as $k_{a}\hat{x}=\eta_{a}(\hat{b}+\hat{b}^{\dagger})$
($a=1,2$). The Lamb--Dicke parameters $\eta_{a}$ of the corresponding
transitions $|3\rangle\rightarrow|a\rangle$ are defined as \begin{equation}
\eta_{a}=k_{a}\Delta x_{0}\qquad(a=1,2).\label{eq:LD-def}\end{equation}
 where $\Delta x_{0}=\sqrt{\hbar/(2m\nu)}$ is the rms spread of the
position of the atom of mass $m$ in the ground state $|0\rangle$
of the trapping potential.

To proceed, the master equation is transformed into a frame, whose
center-of-mass part oscillates with the trap frequency $\nu$ and
whose electronic part rotates with the laser frequency $\omega$.
The transformed master equation is written in terms of three super
operators,\begin{equation}
\frac{\partial\hat{\varrho}}{\partial t}=\left[\hat{\mathcal{N}}+\hat{\mathcal{J}}_{1}(t)+\hat{\mathcal{J}}_{2}(t)\right]\hat{\varrho},\label{eq:master2}\end{equation}
where the evolution in the absence of spontaneous photon emissions
is described by \begin{equation}
\hat{\mathcal{N}}\hat{\varrho}=\frac{1}{i\hbar}\left(\hat{H}_{{\rm eff}}\hat{\varrho}-\hat{\varrho}\hat{H}_{{\rm eff}}^{\dagger}\right).\label{eq:L0}\end{equation}
The occurring non-Hermitean effective Hamiltonian reads \begin{eqnarray}
\hat{H}_{{\rm eff}} & = & \frac{\hbar}{2}\left(\kappa\hat{\sigma}_{1,+}+\kappa^{\ast}\hat{\sigma}_{1,-}\right)+\hbar\Delta|1\rangle\langle1|\nonumber \\
 &  & -\frac{i\hbar}{2}\sum_{a=1,2}\gamma_{a}\hat{\sigma}_{a,+}\hat{\sigma}_{a,-}.\end{eqnarray}
 with $\Delta=\omega-\omega_{1}$ being the detuning of the laser.

The jump operators $\hat{\mathcal{J}}_{a}(t)$ describe spontaneous
photon emissions via the transitions $|3\rangle\to|a\rangle$ ($a=1,2$)
and are given by \begin{equation}
\hat{\mathcal{J}}_{a}(t)\hat{\varrho}=\gamma_{a}\hat{\mathcal{S}}_{a,-}\hat{\mathcal{R}}_{a}(t)\hat{\varrho},\qquad(i=1,2)\label{eq:jump-op}\end{equation}
 where the spin super-operator reads \begin{equation}
\hat{\mathcal{S}}_{a,-}\hat{\varrho}=\hat{\sigma}_{a,-}\hat{\varrho}\hat{\sigma}_{a,+},\label{eq:flip}\end{equation}
 and the now time-dependent recoil operators read \begin{equation}
\hat{\mathcal{R}}_{a}(t)\hat{\varrho}=\int_{-1}^{1}ds\mu_{a}(s)\hat{\mathcal{D}}(i\eta_{a}se^{i\nu t})\hat{\varrho}.\label{eq:r-op}\end{equation}
 The action of the rhs super operator is defined as \begin{equation}
\hat{\mathcal{D}}(\alpha)\hat{\varrho}=\hat{D}(\alpha)\hat{\varrho}\hat{D}^{\dagger}(\alpha),\label{eq:def-sopD}\end{equation}
 with the coherent displacement operator, \begin{equation}
\hat{D}(\alpha)=\exp(\alpha\hat{a}^{\dagger}-\alpha^{\ast}\hat{a}).\label{eq:D-op}\end{equation}
Thus, in Eq. (\ref{eq:r-op}) the length of the shift in motional
phase space depends on both the Lamb--Dicke parameter $\eta_{a}$
and the direction of photon emission $s$, whereas the phase of the
shift is determined by $\pi/2+\nu t$, i.e. it depends on the time
of the spontaneous emission event.

\subsection{Formal solution of the propagator \label{sec:2.2}}

The formal solution of the master equation (\ref{eq:master2}) in
the rotating frame reads \begin{equation}
\hat{\varrho}(t)=\hat{\mathcal{M}}_{012}(t,t_{0})\hat{\varrho}(t_{0}),\label{eq:solution1}\end{equation}
 where $\hat{\varrho}(t_{0})$ is the initial density operator, and
the propagator is given by the time-ordered expression \begin{equation}
\hat{\mathcal{M}}_{012}(t,t_{0})=\mathcal{T}\exp\left\{ \int_{t_{0}}^{t}d\tau\left[\hat{\mathcal{N}}+\hat{\mathcal{J}}_{1}(\tau)+\hat{\mathcal{J}}_{2}(\tau)\right]\right\} .\label{eq:prop012}\end{equation}
 Eq. (\ref{eq:prop012}) can be rewritten as a sum over all possible
trajectories where the spontaneous emissions $|3\rangle\rightarrow|2\rangle$,
i.e. the operator $\hat{\mathcal{J}}_{2}(t)$, occurs at the random
times $t_{1},\ldots,t_{n}$ %
\footnote{This is a standard expansion, known in quantum optics by the term
\char`\"{}unraveling\char`\"{} of the master equation.%
}: \begin{widetext}\begin{equation}
\hat{\mathcal{M}}_{012}(t,t_{0})=\sum_{n=0}^{\infty}\int_{t_{0}}^{t}dt_{n}\ldots\int_{t_{0}}^{t_{2}}dt_{1}\hat{\mathcal{M}}_{01}(t,t_{n})\hat{\mathcal{J}}_{2}(t_{n})\hat{\mathcal{M}}_{01}(t_{n},t_{n-1})\ldots\hat{\mathcal{J}}_{2}(t_{1})\hat{\mathcal{M}}_{01}(t_{1},t_{0}).\label{eq:prop012b}\end{equation}
\end{widetext}The operators $\hat{\mathcal{M}}_{01}(t,t')$ determine
the time evolution between the spontaneous emissions $|3\rangle\rightarrow|2\rangle$
and contain both the laser drive and the spontaneous emissions $|3\rangle\rightarrow|1\rangle$:
\begin{equation}
\hat{\mathcal{M}}_{01}(t,t')=\mathcal{T}\exp\left\{ \int_{t'}^{t}d\tau\left[\hat{\mathcal{N}}+\hat{\mathcal{J}}_{1}(\tau)\right]\right\} .\label{eq:m-operator}\end{equation}

After a spontaneous emission $|3\rangle\rightarrow|2\rangle$ at a
time $t_{k}$, the atom is in its final (pumped) electronic state
$|2\rangle$ where it decouples from the laser drive, and thus $\hat{\mathcal{M}}_{01}(t_{k+1},t_{k})\hat{\mathcal{J}}_{2}(t_{k})=\hat{\mathcal{J}}_{2}(t_{k})$.
Since $\hat{\mathcal{J}}_{2}(t_{k+1})\hat{\mathcal{J}}_{2}(t_{k})=0$
a second spontaneous emission is impossible --- it would require the
electronic state to be $|3\rangle$. Accordingly the propagator (\ref{eq:prop012b})
reduces to only two parts, \begin{equation}
\hat{\mathcal{M}}_{012}(t,t_{0})=\hat{\mathcal{M}}_{01}(t,t_{0})+\int_{t_{0}}^{t}dt^{\prime}\hat{\mathcal{J}}_{2}(t^{\prime})\hat{\mathcal{M}}_{01}(t',t_{0}).\label{eq:solution2}\end{equation}
The first term on the rhs in Eq. (\ref{eq:solution2}) corresponds
to the quantum trajectory when no spontaneous transition from state
$|3\rangle$ to $|2\rangle$ occurred. The second term represents
the trajectory with one final spontaneous transition $|3\rangle\rightarrow|2\rangle$.
Only one such spontaneous transition can take place.

Defining now the projection operators into the electronic states $|a\rangle$,
\begin{equation}
\hat{\mathcal{P}}_{a}\hat{\varrho}=\left(\hat{\sigma}_{a,-}\hat{\sigma}_{a,+}\right)\hat{\varrho}\left(\hat{\sigma}_{a,-}\hat{\sigma}_{a,+}\right),\label{eq:elec-projectors}\end{equation}
and using the facts that (i) the projection into the pumped state
$|2\rangle$ is not affected by the dynamics on the transition $|3\rangle\leftrightarrow|1\rangle$,
that is $\hat{\mathcal{P}}_{2}\hat{\mathcal{M}}_{01}(t,t^{\prime})=\hat{\mathcal{P}}_{2}$,
and that (ii) $\hat{\mathcal{P}}_{2}\hat{\mathcal{J}}_{2}(t^{\prime})=\hat{\mathcal{J}}_{2}(t^{\prime})$,
the part of the propagator taking the atom to the pumped electronic
state $|2\rangle$ is obtained from Eq. (\ref{eq:solution2}) as \begin{equation}
\hat{\mathcal{P}}_{2}\hat{\mathcal{M}}_{012}(t,t_{0})=\hat{\mathcal{P}}_{2}+\int_{t_{0}}^{t}dt^{\prime}\hat{\mathcal{J}}_{2}(t^{\prime})\hat{\mathcal{M}}_{01}(t^{\prime},t_{0}).\label{eq:solution2b}\end{equation}
The first term represents the initial population in level $|2\rangle$,
and the second one the effect of the pumping.

Analogous to Eq. (\ref{eq:prop012b}) the propagator $\hat{\mathcal{M}}_{01}(t,t^{\prime})$
can be formally rewritten as a sum of sequences of spontaneous processes
$|3\rangle\rightarrow|1\rangle$. Denoting by $t_{n}=t^{\prime}$
the time of the final spontaneous emission that leads into the electronic
state $|2\rangle$, Eq. (\ref{eq:solution2b}) thus can be rewritten
as \begin{widetext} \begin{equation}
\hat{\mathcal{P}}_{2}\hat{\mathcal{M}}_{012}(t,t_{0})=\hat{\mathcal{P}}_{2}+\sum_{n=1}^{\infty}\int_{t_{0}}^{t}dt_{n}\ldots\int_{t_{0}}^{t_{2}}dt_{1}\hat{\mathcal{J}}_{2}(t_{n})\hat{\mathcal{M}}_{0}(t_{n},t_{n-1})\hat{\mathcal{J}}_{1}(t_{n-1})\ldots\hat{\mathcal{J}}_{1}(t_{1})\hat{\mathcal{M}}_{0}(t_{1},t_{0}).\label{eq:prop012c}\end{equation}
\end{widetext}Here the operator $\hat{\mathcal{M}}_{0}(t,t^{\prime})$
describes the evolution exclusive of any spontaneous emission, \begin{equation}
\hat{\mathcal{M}}_{0}(t,t^{\prime})=\exp\left[\hat{\mathcal{N}}(t-t^{\prime})\right].\label{eq:N}\end{equation}

A non-vanishing action of $\hat{\mathcal{J}}_{1}(t_{1})$ requires
population in level $|3\rangle$. Since $\hat{\mathcal{M}}_{0}(t_{1},t_{0})$
does only couple levels $|1\rangle$ and $|3\rangle$, the second
term in Eq. (\ref{eq:prop012c}) only acts on populations in levels
$|1\rangle$ and $|3\rangle$. Assuming zero initial population of
$\hat{\varrho}(t_{0})$ in the rapidly decaying state $|3\rangle$,
the action is restricted to population in level $|1\rangle$, and
we may thus multiply the second term in Eq. (\ref{eq:prop012c}) on
its rhs with the projector $\hat{\mathcal{P}}_{1}$, \begin{widetext}
\begin{equation}
\hat{\mathcal{P}}_{2}\hat{\mathcal{M}}_{012}(t,t_{0})=\hat{\mathcal{P}}_{2}+\sum_{n=1}^{\infty}\int_{t_{0}}^{t}dt_{n}\ldots\int_{t_{0}}^{t_{2}}dt_{1}\hat{\mathcal{J}}_{2}(t_{n})\hat{\mathcal{M}}_{0}(t_{n},t_{n-1})\hat{\mathcal{J}}_{1}(t_{n-1})\ldots\hat{\mathcal{M}}_{0}(t_{1},t_{0})\hat{\mathcal{P}}_{1}.\label{eq:prop012d}\end{equation}

\end{widetext}

The second term describes sequences of $n$ consecutive spontaneous
transitions. The laser interaction is restricted to $n$ excitation
processes from level $|1\rangle$ to level $|3\rangle$, \textcolor{black}{as
shown by the sequences of operators $\hat{\mathcal{M}}_{0}$ alternating
with jump operators} $\hat{\mathcal{J}}_{1}\propto\hat{\mathcal{S}}_{1,-}=|1\rangle\langle3|\ldots|3\rangle\langle1|$\textcolor{black}{,
see Eqs (\ref{eq:sigma-a-mp}), (\ref{eq:sigma-a-z}), and (\ref{eq:flip}).
Each of these} excitation processes results therefore in the waiting-time
distribution \cite{wtd-cohen-tannoudji,wtd-zoller,wtd-porrati,wtd-kim}\begin{eqnarray}
w(t-t^{\prime}) & = & (\gamma_{1}+\gamma_{2})\langle3|\left[\hat{\mathcal{M}}_{0}(t,t^{\prime})|1\rangle\langle1|\right]|3\rangle\nonumber \\
 & = & (\gamma_{1}+\gamma_{2})\left|\langle3|\hat{U}_{{\rm eff}}(t-t^{\prime})|1\rangle\right|^{2},\label{eq:w-def}\end{eqnarray}
obtained as the probability (density) of the atom being in level $|3\rangle$
at time $t$, on the condition that it was in $|1\rangle$ at time
$t^{\prime}$, and that no photons have been emitted in the interval
$[t^{\prime},t]$. Here we have defined $\hat{U}_{{\rm eff}}(t)=\exp(-i\hat{H}_{{\rm eff}}t/\hbar)$.
The waiting-time distribution is calculated for resonant pumping in
App. \ref{sec:waiting-time-distribution}.  Using definition (\ref{eq:w-def})
the formal solution (\ref{eq:prop012d}) can therefore be written
as \begin{eqnarray}
\hat{\mathcal{P}}_{2}\hat{\mathcal{M}}_{012}(t,t_{0}) & = & \hat{\mathcal{P}}_{2}+\sum_{n=1}^{\infty}\int_{t_{0}}^{t}dt_{n}\ldots\int_{t_{0}}^{t_{2}}dt_{1}\nonumber \\
 &  & \times[\lambda_{2}\hat{\mathcal{R}}_{2}(t_{n})w(t_{n}-t_{n-1})]\nonumber \\
 &  & \times\left\{ [\lambda_{1}\hat{\mathcal{R}}_{1}(t_{n-1})w(t_{n-1}-t_{n-2})]\ldots\right.\nonumber \\
 &  & \left.\quad\ldots[\lambda_{1}\hat{\mathcal{R}}_{1}(t_{1})w(t_{1}-t_{0})]\right\} \hat{\mathcal{P}}_{21}\label{eq:sol-expl}\end{eqnarray}
where \begin{equation}
\lambda_{a}=\gamma_{a}/(\gamma_{1}+\gamma_{2}),\qquad(a=1,2)\label{eq:branching-ratios}\end{equation}
are the branching ratios, and the operator that induces the transition
into the final level $|2\rangle$ is defined as\begin{equation}
\hat{\mathcal{P}}_{21}\hat{\varrho}=\left(\hat{\sigma}_{2,-}\hat{\sigma}_{1,+}\right)\hat{\varrho}\left(\hat{\sigma}_{1,-}\hat{\sigma}_{2,+}\right).\label{eq:def-projectorP21}\end{equation}

Defining  the joint probability density for $n$ spontaneous transitions
at times $t_{1},\ldots,t_{n}$ as \begin{eqnarray}
w_{n}(t_{n},\ldots,t_{1}) & = & \lambda_{2}\lambda_{1}^{n-1}w\left(t_{n}-t_{n-1}\right)\ldots\label{eq:photon-joint-prob}\\
 &  & \quad\ldots w\left(t_{2}-t_{1}\right)w\left(t_{1}-t_{0}\right),\quad(n>0)\nonumber \end{eqnarray}
the result (\ref{eq:sol-expl}) is rewritten as \begin{widetext}

\begin{equation}
\hat{\mathcal{P}}_{2}\hat{\mathcal{M}}_{012}(t,t_{0})=\hat{\mathcal{P}}_{2}+\sum_{n=1}^{\infty}\int_{t_{0}}^{t}dt_{n}\ldots\int_{t_{0}}^{t_{2}}dt_{1}w_{n}\left(t_{n},\ldots,t_{1}\right)\hat{\mathcal{R}}_{2}(t_{n})\hat{\mathcal{R}}_{1}(t_{n-1})\ldots\hat{\mathcal{R}}_{1}(t_{1})\hat{\mathcal{P}}_{21}.\label{eq:sol-expl2}\end{equation}
\end{widetext}

\section{The Vibrational Effects Of The Optical Transitions \label{sec:3}}

In Eq. (\ref{eq:sol-expl2}), a sequence of recoil operators, cf.
Eq. (\ref{eq:r-op}), appears, which can be combined as\begin{eqnarray}
\hat{\mathcal{R}}_{2}(t_{n})\ldots\hat{\mathcal{R}}_{1}(t_{1}) & = & \int ds_{n}\mu_{2}(s_{n})\ldots\int ds_{1}\mu_{1}(s_{1})\nonumber \\
 & \times & \hat{\mathcal{D}}(i\eta_{2}s_{n}e^{i\nu t_{n}})\hat{\mathcal{D}}(i\eta_{1}s_{n-1}e^{i\nu t_{n-1}})\ldots\nonumber \\
 &  & \ldots\hat{\mathcal{D}}(i\eta_{1}s_{1}e^{i\nu t_{1}})\nonumber \\
 & = & \int ds_{n}\mu_{2}(s_{n})\ldots\int ds_{1}\mu_{1}(s_{1})\nonumber \\
 &  & \times\hat{\mathcal{D}}\left[\alpha_{n}(\{ t_{n},s_{n}\},\ldots,\{ t_{1},s_{1}\})\right],\label{eq:shift-combined}\end{eqnarray}
where the total recoil shift is\begin{equation}
\alpha_{n}(\{ t_{n},s_{n}\},\ldots,\{ t_{1},s_{1}\})=i\eta_{2}s_{n}e^{i\nu t_{n}}+\sum_{m=1}^{n-1}i\eta_{1}s_{m}e^{i\nu t_{m}}.\label{eq:disp-traj}\end{equation}
The $n$-fold integration represents the averaging over the directions
of recoil in the $n$ consecutive spontaneous emissions at times $t_{1},\ldots,t_{n}$.
In order to consider any number of spontaneous emissions at arbitrary
moments of time within the interval $[t_{0},t]$, we must average
over the joint photon-emission probability at the consecutive emission
times. In this way we obtain the operator of the mean recoil shift
in phase space: \begin{widetext} \begin{eqnarray}
\hat{\overline{\mathcal{D}}}(t,t_{0}) & = & \sum_{n=1}^{\infty}\int_{t_{0}}^{t}dt_{n}\ldots\int_{t_{0}}^{t_{2}}dt_{1}w_{n}(t_{n},\ldots,t_{1})\hat{\mathcal{R}}_{2}(t_{n})\hat{\mathcal{R}}_{1}(t_{n-1})\ldots\hat{\mathcal{R}}_{1}(t_{1})\nonumber \\
 & = & \sum_{n=1}^{\infty}\int_{t_{0}}^{t}dt_{n}\int ds_{n}\mu_{2}(s_{n})\ldots\int_{t_{0}}^{t_{2}}dt_{1}\int ds_{1}\mu_{1}(s_{1})w_{n}(t_{n},\ldots,t_{1})\nonumber \\
 &  & \quad\times\hat{\mathcal{D}}\left[\alpha_{n}(\{ t_{n},s_{n}\},\ldots,\{ t_{1},s_{1}\})\right].\label{eq:avg-recoil}\end{eqnarray}

\end{widetext}

One may define the density of recoil shifts in phase-space that contains
all the recoil effects of the spontaneous emissions at random times
into random directions, \begin{eqnarray}
p(\alpha;t,t_{0}) & = & \sum_{n=1}^{\infty}\int_{t_{0}}^{t}dt_{n}\int ds_{n}\ldots\int_{t_{0}}^{t_{2}}dt_{1}\int ds_{1}\nonumber \\
 & \times & \mu_{2}(s_{n})\mu_{1}(s_{n-1})\ldots\mu_{1}(s_{1})w_{n}(t_{n},\ldots,t_{1})\nonumber \\
 & \times & \delta\left[\alpha-\alpha_{n}(\{ t_{n},s_{n}\},\ldots,\{ t_{1},s_{1}\})\right].\label{eq:displ-distr}\end{eqnarray}
The operator of the mean recoil shift in phase space (\ref{eq:avg-recoil})
is expressed as an integral over $\hat{\mathcal{D}}$ weighted by
this distribution $p(\alpha;t,t_{0})$ in phase-space:\begin{equation}
\hat{\overline{\mathcal{D}}}(t,t_{0})=\int d^{2}\alpha p(\alpha;t,t_{0})\hat{\mathcal{D}}(\alpha).\label{eq:avg-recoil2}\end{equation}
Since $p(\alpha;t,t_{0})$ is an integral over positive probabilities,
it may be viewed as a blurring distribution that smoothes the initial
vibrational phase-space distribution. It contains all the characteristics
of the random process that acts on the center-of-mass degree of freedom
of the electronic population that is transferred from state $|1\rangle$
to state $|2\rangle$.

In the propagator of Eq. (\ref{eq:sol-expl2}), leading to the pumped
electronic state $|2\rangle$, the second term on the rhs contains
the averaged displacement as given in Eq. (\ref{eq:avg-recoil}),
and thus Eq. (\ref{eq:avg-recoil2}) may be employed to yield

\begin{equation}
\hat{\mathcal{P}}_{2}\hat{\mathcal{M}}_{012}(t,t_{0})=\hat{\mathcal{P}}_{2}+\int d^{2}\alpha p(\alpha;t,t_{0})\hat{\mathcal{D}}(\alpha)\hat{\mathcal{P}}_{21}.\label{eq:solution7}\end{equation}

Applying Eq. (\ref{eq:solution7}) on the initial vibronic density
operator $\hat{\varrho}(t_{0})$ and tracing over the electronic degree
of freedom, we obtain the reduced motional density operator for the
population in the pumped electronic state,\begin{equation}
\langle2|\hat{\varrho}(t)|2\rangle=\langle2|\hat{\varrho}(t_{0})|2\rangle+\int d^{2}\alpha p(\alpha;t,t_{0})\hat{\mathcal{D}}(\alpha)\langle1|\hat{\varrho}(t_{0})|1\rangle.\label{eq:solution7b}\end{equation}
A vibrational density operator $\langle a|\hat{\varrho}|a\rangle$,
that corresponds to the population in electronic level $|a\rangle$,
is related to the $s$-ordered phase-space distribution \begin{equation}
P_{a}^{(s)}(\alpha)=\frac{1}{P_{a}}{\rm Tr}\left[\hat{\delta}^{(s)}(\alpha-\hat{b})\langle a|\hat{\varrho}|a\rangle\right],\qquad(a=1,2),\label{eq:wigner}\end{equation}
where the $s$-ordered delta operator is formally defined via the
displacement operator (\ref{eq:D-op}) as \cite{glauber1,glauber2}
\begin{equation}
\hat{\delta}^{(s)}(\alpha-\hat{b})=\frac{1}{\pi^{2}}\int d^{2}\beta\exp\left(\alpha\beta^{\ast}-\beta\alpha^{\ast}+s|\beta|^{2}/2\right)\hat{D}(\beta),\label{eq:delta-op}\end{equation}
and the probability to find the atom in the electronic state $|a\rangle$
is defined by \begin{equation}
P_{a}={\rm Tr}\left(\langle a|\hat{\varrho}|a\rangle\right).\label{eq:prob-a-def}\end{equation}

We want to finally come up with a mapping of the initial vibrational
phase-space distribution on the distribution generated by the optical
pumping process. For this purpose we apply Eq. (\ref{eq:wigner})
on the mapping of vibrational density operators, Eq. (\ref{eq:solution7b}),
making use of the definition (\ref{eq:def-sopD}) and obtain\begin{eqnarray}
P_{2}^{(s)}(\alpha,t) & = & P_{2}(t_{0})P_{2}^{(s)}(\alpha,t_{0})+\int d^{2}\alpha^{\prime}p(\alpha^{\prime};t,t_{0})\label{eq:wigner-map1}\\
 &  & \quad\times{\rm Tr}\left[\hat{\delta}^{(s)}(\alpha-\hat{b})\hat{D}(\alpha^{\prime})\langle1|\hat{\varrho}(t_{0})|1\rangle\hat{D}^{\dagger}(\alpha^{\prime})\right].\nonumber \end{eqnarray}
Employing the cyclic property of the trace, one rewrites the trace
as 

\begin{eqnarray}
 &  & {\rm Tr}\left[\hat{D}^{\dagger}(\alpha^{\prime})\hat{\delta}^{(s)}(\alpha-\hat{b})\hat{D}(\alpha^{\prime})\langle1|\hat{\varrho}(t_{0})|1\rangle\right]\nonumber \\
 &  & ={\rm Tr}\left[\hat{\delta}^{(s)}\left[\alpha-(\hat{b}+\alpha^{\prime})\right]\langle1|\hat{\varrho}(t_{0})|1\rangle\right],\label{eq:cyclic-property}\end{eqnarray}
where the displacement operation results in a shift of the vibrational
operator $\hat{b}\to\hat{b}+\alpha^{\prime}$. This result, with substitution
of integration variable $\alpha^{\prime}\to\alpha-\alpha^{\prime}$
in Eq. (\ref{eq:wigner-map1}), yields\begin{eqnarray}
P_{2}^{(s)}(\alpha,t) & = & P_{2}(t_{0})P_{2}^{(s)}(\alpha,t_{0})+\int d^{2}\alpha^{\prime}p(\alpha-\alpha^{\prime};t,t_{0})\nonumber \\
 &  & \quad\times{\rm Tr}\left[\hat{\delta}^{(s)}\left[\alpha^{\prime}-\hat{b}\right]\langle1|\hat{\varrho}(t_{0})|1\rangle\right].\label{eq:wigner-map2}\end{eqnarray}
With the use of definition (\ref{eq:wigner}) this phase-space distribution
writes \begin{eqnarray}
P_{2}^{(s)}(\alpha,t) & = & P_{2}(t_{0})P_{2}^{(s)}(\alpha,t_{0})\label{eq:map}\\
 &  & +P_{1}(t_{0})\int d^{2}\alpha^{\prime}p(\alpha-\alpha^{\prime};t,t_{0})P_{1}^{(s)}(\alpha^{\prime},t_{0}).\nonumber \end{eqnarray}
\textcolor{black}{Thus the initial phase-space distribution of the
population in electronic level $|1\rangle$ is mapped onto a contribution
to the phase-space distribution of the pumped level $|2\rangle$ via
a convolution with the density of recoil shifts in phase-space (\ref{eq:displ-distr}).}

\section{The Phase Space Density of Recoil Shifts\label{sec:4}}

\subsection{Moments of the density of recoil shifts}

The density $p$ of recoil shifts in phase space multiplied by an
infinitesimal phase-space volume, $p(\alpha;t,t_{0})d^{2}\alpha$,
is the probability of displacing, by $\alpha$, the initial phase-space
distribution within the time interval $[t_{0},t]$. The observable
features of the collective recoil shift will be characterized by this
density's statistical moments of all orders $(k,l)$,\begin{equation}
\langle\alpha^{\ast k}\alpha^{l}\rangle_{p(\alpha;t,t_{0})}=\int d^{2}\alpha\alpha^{\ast k}\alpha^{l}p(\alpha;t,t_{0}).\label{eq:B-moments}\end{equation}
To evaluate these moments we may employ the characteristic function
$\underline{p}(\alpha,t,t_{0})$, which is the Fourier transform of
the density of recoil shifts,\begin{equation}
\underline{p}(\alpha;t,t_{0})=\int d^{2}\beta\exp(\alpha\beta^{\ast}-\beta\alpha^{\ast})p(\beta;t,t_{0}).\label{eq:charac0}\end{equation}
With the help of this characteristic function the statistical moments,
Eq. (\ref{eq:B-moments}), are obtained as \cite{gardiner}

\begin{equation}
\langle\alpha^{\ast k}\alpha^{l}\rangle_{p(\alpha;t,t_{0})}=(-1)^{l}\left[\partial_{\alpha}^{k}\partial_{\alpha^{\ast}}^{l}\underline{p}(\alpha,t,t_{0})\right]_{\alpha=0}.\label{eq:P-moments}\end{equation}

Inserting Eq. (\ref{eq:displ-distr}) into (\ref{eq:charac0}) and
performing the integral over $\beta$ yields 

\begin{eqnarray}
\underline{p}(\alpha;t,t_{0}) & = & \sum_{n=1}^{\infty}\int_{t_{0}}^{t}dt_{n}\int ds_{n}\ldots\int_{t_{0}}^{t_{2}}dt_{1}\int ds_{1}\label{eq:charac1}\\
 & \times & \mu_{2}(s_{n})\mu_{1}(s_{n-1})\ldots\mu(s_{1})w_{n}(t_{n},\ldots,t_{1})\nonumber \\
 & \times & \exp\left[\alpha\alpha_{n}^{\ast}(\{ t_{n},s_{n}\},\ldots,\{ t_{1},s_{1}\})-{\rm c.c.}\right],\nonumber \end{eqnarray}
where $\alpha_{n}(\{ t_{n},s_{n}\},\ldots,\{ t_{1},s_{1}\})$ is defined
in Eq. (\ref{eq:disp-traj}). With this definition, the exponential
factor writes\begin{eqnarray}
 &  & \exp\left[\alpha\alpha_{n}^{\ast}(\{ t_{n},s_{n}\},\ldots,\{ t_{1},s_{1}\})-{\rm c.c.}\right]\nonumber \\
 &  & =\exp\left[\alpha\left(i\eta_{2}s_{n}e^{i\nu t_{n}}+\sum_{m=1}^{n-1}i\eta_{1}s_{m}e^{i\nu t_{m}}\right)^{\ast}-{\rm c.c.}\right]\nonumber \\
 &  & =\exp\left(-\alpha i\eta_{2}s_{n}e^{-i\nu t_{n}}-{\rm c.c.}\right)\nonumber \\
 &  & \quad\times\prod_{m=1}^{n-1}\exp\left(-\alpha i\eta_{1}s_{m}e^{-i\nu t_{m}}-{\rm c.c.}\right).\label{eq:exp-factor-simplify}\end{eqnarray}
Each exponential factor on the rhs can be rewritten as\begin{eqnarray}
 &  & \exp\left(-\alpha i\eta se^{-i\nu t}-{\rm c.c.}\right)\nonumber \\
 &  & =\exp\left[-2i\eta s|\alpha|\Re\left(e^{i\nu t-i\arg\alpha}\right)\right]\nonumber \\
 &  & =\exp\left[-2i\eta s|\alpha|\cos(\nu t-\arg\alpha)\right],\label{eq:exp-factor-simplify2}\end{eqnarray}
so that Eq. (\ref{eq:charac1}) reads \begin{widetext}\begin{eqnarray}
\underline{p}(\alpha;t,t_{0}) & = & \sum_{n=1}^{\infty}\int_{t_{0}}^{t}dt_{n}\ldots\int_{t_{0}}^{t_{2}}dt_{1}w_{n}(t_{n},\ldots,t_{1})\int ds_{n}\mu_{2}(s_{n})e^{-2i\eta_{2}s_{n}|\alpha|\cos(\nu t_{n}-\arg\alpha)}\label{eq:B-charac}\\
 &  & \times\int ds_{n-1}\mu_{1}(s_{n-1})e^{-2i\eta_{1}s_{n-1}|\alpha|\cos(\nu t_{n-1}-\arg\alpha)}\ldots\int ds_{1}\mu_{1}(s_{1})e^{-2i\eta_{1}s_{1}|\alpha|\cos(\nu t_{1}-\arg\alpha)}.\nonumber \end{eqnarray}
\end{widetext}Each integral over a dipole characteristics is a directional
spectrum of recoils; it is formally expressed by the Fourier transform\begin{equation}
\underline{\mu}_{a}(\xi)=\int_{-1}^{1}ds\mu_{a}(s)\exp(-is\xi).\label{eq:mubar-def}\end{equation}
In Eq. (\ref{eq:B-charac}) this directional spectrum appears $n$
times with the arguments $2\eta_{a}|\alpha|\cos(\nu t_{m}-\arg\alpha)$,
where $m=1,\ldots,n$, and $a=1,2$ depending on the decay channel.
Each of these spectra defines a characteristic function associated
to the instantaneous probability density for a single recoil shift
at time $t$, \begin{equation}
\underline{p}_{a}(\alpha,t)=\underline{\mu}_{a}\left[2\eta_{a}|\alpha|\cos(\nu t-\arg\alpha)\right],\qquad(a=1,2).\label{eq:tilde-mu2}\end{equation}
With these replacements Eq. (\ref{eq:B-charac}) takes on the form

\begin{eqnarray}
\underline{p}(\alpha;t,t_{0}) & = & \sum_{n=1}^{\infty}\int_{t_{0}}^{t}dt_{n}\ldots\int_{t_{0}}^{t_{2}}dt_{1}w_{n}(t_{n},\ldots,t_{1})\nonumber \\
 &  & \times\underline{p}_{2}(\alpha,t_{n})\underline{p}_{1}(\alpha,t_{n-1})\ldots\underline{p}_{1}(\alpha,t_{1}).\label{eq:charac2}\end{eqnarray}

The moments of the density of recoil shifts \textcolor{black}{are
evaluated according to Eq. (\ref{eq:P-moments}).} The characteristic
function (\ref{eq:charac2}) depends on $\alpha$ via the factors
$\underline{p}_{a}(\alpha,t)$. The derivatives of these factors \textcolor{black}{(App.
\ref{app:deriv-charac})} are\begin{equation}
\left[\partial_{\alpha}^{k}\partial_{\alpha^{\ast}}^{l}\underline{p}_{a}(\alpha,t)\right]_{\alpha=0}=(-i\eta_{a})^{k+l}e^{i\nu t(l-k)}\left\langle (\cos\theta)^{k+l}\right\rangle _{a}\label{eq:mu-derivs}\end{equation}
with $a=1,2$. The moments of the projection of the spontaneously
emitted photon's wave vector on the dipole axis are \begin{equation}
\left\langle (\cos\theta)^{k}\right\rangle _{a}=\int_{-1}^{1}dss^{k}\mu_{a}(s).\label{eq:def-muk}\end{equation}
Since the dipole radiation characteristics shows even spatial symmetry,
$\mu_{a}(-s)=\mu_{a}(s)$, cf. Eq. (\ref{eq:dip-charac-app2}), all
odd moments vanish, $\langle(\cos\theta)^{2k+1}\rangle_{a}=0$ ($k=0,1,2,\ldots$).
In consequence all odd-order derivatives of $\underline{p}_{a}(\beta,t)$
vanish, see Eq. (\ref{eq:mu-derivs}), and via Eq. (\ref{eq:charac2})
the odd-order moments of $p$ vanish likewise: \begin{equation}
\langle\alpha^{\ast k}\alpha^{l}\rangle_{p(\alpha;t,t_{0})}=0\quad\mbox{if $k+l=2n+1$}.\label{eq:odd}\end{equation}
This result could have been anticipated, since the incoherent scattering
process connected with optical pumping cannot generate a coherent
vibrational amplitude.

The non-vanishing \emph{even} moments of a complete optical-pumping
process ($t\to\infty)$ are \begin{eqnarray}
\langle\alpha^{\ast k}\alpha^{l}\rangle_{p}=(-1)^{l}\sum_{n=1}^{\infty}\int_{t_{0}}^{\infty}dt_{n}\ldots\int_{t_{0}}^{t_{2}}dt_{1}w_{n}(t_{n},\ldots,t_{1})\nonumber \\
\quad\times\left\{ \partial_{\alpha}^{k}\partial_{\alpha^{\ast}}^{l}\left[\underline{p}_{2}(\alpha,t_{n})\underline{p}_{1}(\alpha,t_{n-1})\ldots\underline{p}_{1}(\alpha,t_{1})\right]\right\} _{\alpha=0}.\quad\label{eq:P-moments3}\end{eqnarray}
For the lowest-order non-trivial even moment $\langle\alpha^{\ast}\alpha\rangle_{p}$
two derivatives have to be performed on any same factor $\underline{p}_{a}(\alpha,t)$
which results in a sum of $n$ terms, where the first one reads\begin{eqnarray}
 &  & -\sum_{n=1}^{\infty}\int_{t_{0}}^{\infty}dt_{n}\ldots\int_{t_{0}}^{t_{2}}dt_{1}w_{n}(t_{n},\ldots,t_{1})\label{eq:first-one1}\\
 &  & \quad\times\left\{ \left[\partial_{\alpha}\partial_{\alpha^{\ast}}\underline{p}_{2}(\alpha,t_{n})\right]\underline{p}_{1}(\alpha,t_{n-1})\ldots\underline{p}_{1}(\alpha,t_{1})\right\} _{\alpha=0}.\nonumber \end{eqnarray}
Given that $\underline{p}_{a}(0,t)=\underline{\mu}_{a}(0)=1$ {[}Eqs
(\ref{eq:mubar-def}) and (\ref{eq:tilde-mu2})] and $[\partial_{\alpha}\partial_{\alpha^{\ast}}\underline{p}_{2}(\alpha,t_{n})]_{\alpha=0}=-\eta_{2}^{2}\langle(\cos\theta)^{2}\rangle_{2}$,
the above term is\begin{eqnarray}
 &  & \eta_{2}^{2}\langle(\cos\theta)^{2}\rangle_{2}\sum_{n=1}^{\infty}\int_{t_{0}}^{\infty}dt_{n}\ldots\int_{t_{0}}^{t_{2}}dt_{1}w_{n}(t_{n},\ldots,t_{1})\nonumber \\
 &  & =\eta_{2}^{2}\langle(\cos\theta)^{2}\rangle_{2}\sum_{n=1}^{\infty}P_{n}\nonumber \\
 &  & =\eta_{2}^{2}\langle(\cos\theta)^{2}\rangle_{2}.\label{eq:first-one1-simplify}\end{eqnarray}
Here the statistics of the number of spontaneously emitted photons
during the pump process is introduced as time-ordered integral of
the joint probability density of photon emissions,\begin{equation}
P_{n}=\int_{t_{0}}^{\infty}dt_{n}\ldots\int_{t_{0}}^{t_{2}}dt_{1}w_{n}(t_{n},\ldots,t_{1}),\quad(n\geq1),\label{eq:photon-prob-def-first}\end{equation}
and $P_{0}=0$. The moments of this photon statistics are evaluated
in App. \ref{app:photon-stat}.

The sum of the other $n-1$ terms is

\begin{eqnarray}
 &  & -\sum_{n=1}^{\infty}\int_{t_{0}}^{\infty}dt_{n}\ldots\int_{t_{0}}^{t_{2}}dt_{1}w_{n}(t_{n},\ldots,t_{1}\})\sum_{p=1}^{n-1}\left\{ \underline{p}_{2}(\alpha,t_{n})\right.\nonumber \\
 &  & \quad\left.\times\underline{p}_{1}(\alpha,t_{n-1})\ldots\left[\partial_{\alpha}\partial_{\alpha^{\ast}}\underline{p}_{1}(\alpha,t_{p})\right]\ldots\underline{p}_{1}(\alpha,t_{1})\right\} _{\alpha=0}.\qquad\label{eq:sum-of-the-other}\end{eqnarray}
As the differentiated factor is $-\eta_{1}^{2}\langle(\cos\theta)^{2}\rangle_{1}$
in all $n-1$ terms, they sum up to yield\begin{eqnarray}
 &  & \eta_{1}^{2}\langle(\cos\theta)^{2}\rangle_{1}\sum_{n=1}^{\infty}\int_{t_{0}}^{\infty}dt_{n}\ldots\int_{t_{0}}^{t_{2}}dt_{1}w_{n}(t_{n},\ldots,t_{1})(n-1)\nonumber \\
 &  & =\eta_{1}^{2}\langle(\cos\theta)^{2}\rangle_{1}\sum_{n=1}^{\infty}P_{n}(n-1)\nonumber \\
 &  & =\eta_{1}^{2}\langle(\cos\theta)^{2}\rangle_{1}\left(\langle\hat{n}_{{\rm ph}}\rangle-1\right),\label{eq:differentiated-factor}\end{eqnarray}
where the mean number of emitted photons is given in Eq. (\ref{eq:n-avg}).
Thus, the lowest-order non-trivial moment of the density of recoil
shifts is \begin{equation}
\langle\alpha^{\ast}\alpha\rangle_{p}=\eta_{2}^{2}\langle(\cos\theta)^{2}\rangle_{2}+\eta_{1}^{2}(\langle\hat{n}_{{\rm ph}}\rangle-1)\langle(\cos\theta)^{2}\rangle_{1}.\label{eq:n-avg2}\end{equation}
Upon insertion of the mean number of spontaneously emitted photons,
see (\ref{eq:n-avg}), this moment becomes\begin{equation}
\bar{n}_{p}=\langle\alpha^{\ast}\alpha\rangle_{p}=\eta_{2}^{2}\langle(\cos\theta)^{2}\rangle_{2}+\frac{\lambda_{1}}{\lambda_{2}}\eta_{1}^{2}\langle(\cos\theta)^{2}\rangle_{1},\label{eq:n-avg2b}\end{equation}
where the mean recoil-induced excitation $\bar{n}_{p}=\langle\alpha^{\ast}\alpha\rangle_{p}$
is introduced. This moment is the width of the density distribution
of recoil shifts to which the system has been subjected in the optical
pumping process. 

In the same order one obtains the moment $\langle\alpha^{2}\rangle_{p}$
which gives information on the rotational symmetry of the density
of recoil shifts in phase space. Analogously to the above calculation
one obtains this moment -- see App. \ref{app:a2mom} -- as\begin{eqnarray}
\langle\alpha^{2}\rangle_{p} & = & -\left(\eta_{2}^{2}\langle(\cos\theta)^{2}\rangle_{2}+\frac{\lambda_{1}}{\lambda_{2}}\eta_{1}^{2}\langle(\cos\theta)^{2}\rangle_{1}\right)\nonumber \\
 &  & \quad\times\,\frac{\lambda_{2}\underline{w}(2\nu)}{1-\lambda_{1}\underline{w}(2\nu)},\label{eq:a2mom-result}\end{eqnarray}
where $t_{0}=0$ was set to cancel the trivial phase factor $e^{2i\nu t_{0}}$,
and the spectral waiting-time distribution is defined as\begin{equation}
\underline{w}(\omega)=\int_{0}^{\infty}dtw(t)e^{i\omega t}.\label{eq:spectral-wait-time-def}\end{equation}
with $w(t)$ defined in Eq. (\ref{eq:w-def}).

The last complex-valued factor in Eq. (\ref{eq:a2mom-result}), \begin{equation}
\frac{\lambda_{2}\underline{w}(2\nu)}{1-\lambda_{1}\underline{w}(2\nu)}=Ae^{i\phi_{A}},\label{eq:coef-def}\end{equation}
-- where $A$ is a real-valued semi-positive number -- depends on
the branching ratio, and, via the waiting-time distribution, on the
trap frequency and the laser parameters. This function characterizes
the saturation of the recoil distribution's anisotropy. With the definition
\begin{equation}
\underline{w}(2\nu)=\underline{w}\exp(i\phi_{\underline{w}}),\label{eq:w-polar}\end{equation}
and since $\underline{w}$ is semi-positive, the modulus $A$ and
the phase $\phi_{A}$ of the anisotropy (\ref{eq:coef-def}) can be
written as\begin{eqnarray}
A & = & \frac{\lambda_{2}\underline{w}}{\sqrt{(1-\lambda_{1}\underline{w})^{2}+4\lambda_{1}\underline{w}\sin^{2}(\phi_{\underline{w}}/2)}},\label{eq:A-modulus}\\
\tan\phi_{A} & = & \frac{\sin\phi_{\underline{w}}}{\cos\phi_{\underline{w}}-\lambda_{1}\underline{w}}.\label{eq:A-phase}\end{eqnarray}
Employing Eqs (\ref{eq:n-avg2b}) and (\ref{eq:coef-def}), we rewrite
the moment (\ref{eq:a2mom-result}) as\begin{equation}
\langle\alpha^{2}\rangle_{p}=-\bar{n}_{p}Ae^{i\phi_{A}}.\label{eq:aa-new}\end{equation}

The phase-dependent quadrature is defined as\begin{equation}
q(\phi)=\frac{1}{\sqrt{2}}\left(\alpha e^{i\phi}+\alpha^{\ast}e^{-i\phi}\right)=\sqrt{2}\Re\left(\alpha e^{i\phi}\right).\label{eq:def-q-class}\end{equation}
The above considerations suggest that the average quadrature vanishes,
$\langle q(\phi)\rangle_{p}=0$. Its variance, however, is in general
non-vanishing and phase dependent. It is expressed in terms of the
moments {[}$\Delta q(\phi)=q(\phi)-\langle q(\phi)\rangle_{p}$] as\begin{equation}
\langle[\Delta q(\phi)]^{2}\rangle_{p}=\langle\alpha^{\ast}\alpha\rangle_{p}+\Re\left(\langle\alpha^{2}\rangle_{p}e^{2i\phi}\right),\label{eq:qq-map0}\end{equation}
which, upon insertion of Eqs. (\ref{eq:n-avg2b}) and (\ref{eq:aa-new}),
writes\begin{equation}
\delta q_{p,\phi}^{2}=\langle[\Delta q(\phi)]^{2}\rangle_{p}=\bar{n}_{p}\left[1-A\cos\left(2\phi+\phi_{A}\right)\right],\label{eq:qq-map}\end{equation}
where the abbreviation $\delta q_{p,\phi}=\sqrt{\langle[\Delta q(\phi)]^{2}\rangle_{p}}$
has been introduced.

At the values of the phase $\phi_{-}=-\phi_{A}/2$ and $\phi_{+}=\phi_{-}+\pi/2$,
the minimum and maximum variances, respectively, are attained,\begin{equation}
\delta q_{p,\phi_{\pm}}^{2}=\bar{n}_{p}(1\pm A),\label{eq:minmax-quadrature}\end{equation}
and the product of the unequal rms uncertainties is\begin{equation}
\delta q_{p,\phi_{+}}\cdot\delta q_{p,\phi_{-}}=\bar{n}_{p}\sqrt{1-A^{2}}.\label{eq:minmax-quad-prod}\end{equation}
In general the density of recoil shifts is not rotationally symmetric,
as shows the ratio of the difference of maximum and minimum quadrature
fluctuations over the mean fluctuation, i.e. the anisotropy in the
quadratures:\begin{equation}
\frac{\delta q_{p,\phi_{+}}^{2}-\delta q_{p,\phi_{-}}^{2}}{\delta q_{p,\phi_{+}}^{2}+\delta q_{p,\phi_{-}}^{2}}=A.\label{eq:def-anisotropy-quad}\end{equation}

There are, however, two special cases where the density is approximately
isotropic, i.e. where $A\approx0$. This situation is obtained either
with \textcolor{black}{negligible} spectral waiting-time distribution
at twice the trap frequency, $\underline{w}=|\underline{w}(2\nu)|\approx0$,
or with a very large value of the branching ratio, $\lambda_{1}\approx1$
and $\lambda_{2}\approx0$, see Eq. (\ref{eq:A-modulus}). The first
case represents a situation where the spectral waiting-time distribution
lacks a frequency component at twice the trap frequency. This condition
means that the waiting times of the probabilistic sequence of spontaneous
emissions of photons are not synchronized to the half period of the
trap oscillation. The latter case simply corresponds to a situation
where the decay rate to state $|1\rangle$ much exceeds that to state
$|2\rangle$, so that the optical pumping process requires a very
long sequence of spontaneous photon emissions, each one randomizing
the density of recoil shifts that eventually becomes isotropic. This
case corresponds to fluorescence scattering, as discussed for a free
atom in Refs \cite{pusep,mandel}. In general, however, a finite sequence
of photon emissions during the optical pumping generates a {}``squashed''
density of recoil shifts with a predefined orientation in phase space
that is determined by the phase of the spectral waiting-time distribution.

Finally, for the characterization of the density of recoil shifts,
the variance \begin{equation}
\delta n_{p}^{2}=\langle\alpha^{\ast2}\alpha^{2}\rangle_{p}-\bar{n}_{p}^{2},\label{eq:np-var}\end{equation}
that corresponds to the mean $\bar{n}_{p}$, is evaluated. The first
moment on the rhs is obtained as {[}see Eq. (\ref{eq:a4-moment})]\begin{eqnarray}
\langle\alpha^{\ast2}\alpha^{2}\rangle_{p} & = & \left(\eta_{2}^{4}\langle(\cos\theta)^{4}\rangle_{2}+\frac{\lambda_{1}}{\lambda_{2}}\eta_{1}^{4}\langle(\cos\theta)^{4}\rangle_{1}\right)\label{eq:a4-moment-def}\\
 & + & 2\bar{n}_{p}\left(\bar{n}_{p}-\eta_{2}^{2}\langle(\cos\theta)^{2}\rangle_{2}\right)\left(1+A\cos\phi_{A}\right),\nonumber \end{eqnarray}
so that the variance (\ref{eq:np-var}) becomes\begin{eqnarray}
\delta n_{p}^{2} & = & \left(\eta_{2}^{4}\langle(\cos\theta)^{4}\rangle_{2}+\frac{\lambda_{1}}{\lambda_{2}}\eta_{1}^{4}\langle(\cos\theta)^{4}\rangle_{1}\right)-\bar{n}_{p}^{2}\nonumber \\
 & + & 2\bar{n}_{p}\left(\bar{n}_{p}-\eta_{2}^{2}\langle(\cos\theta)^{2}\rangle_{2}\right)\left(1+A\cos\phi_{A}\right).\label{eq:np-var2}\end{eqnarray}

\subsection{Mapping the moments of the atomic phase-space distribution\label{sec:3.1}}

So far, the density of recoil shifts has been characterized by its
statistical moments. To characterize the vibrational quantum state
of the ion after completion of optical pumping, the initial vibrational
phase-space distribution of the ion has to be convolved with the density
of recoil shifts, as given in Eq. (\ref{eq:map}). Fourier transforming
Eq. (\ref{eq:map}), one obtains the mapping of the characteristic
function, that corresponds to the vibrational phase-space distribution,
as \begin{equation}
\underline{P}_{2}^{(s)}(\alpha,t)=P_{2}(t_{0})\underline{P}_{2}^{(s)}(\alpha,t_{0})+P_{1}(t_{0})\underline{p}(\alpha;t,t_{0})\underline{P}_{1}^{(s)}(\alpha,t_{0}),\label{eq:map2}\end{equation}
where the Fourier transform of the density of recoil shifts is given
by Eq. (\ref{eq:charac2}).

From the $s$-ordered phase-space distributions, $P_{a}^{(s)}(\alpha,t)$,
the $s$-ordered quantum-statistical moments are represented by integrals
\cite{glauber1,glauber2}, \begin{eqnarray}
\left\langle \left\{ \hat{b}^{\dagger k}(t)\hat{b}^{l}(t)\right\} _{s}\right\rangle _{a} & = & \frac{1}{P_{a}}{\rm Tr}\left[\left\{ \hat{b}^{\dagger k}\hat{b}^{l}\right\} _{s}\langle a|\hat{\varrho}(t)|a\rangle\right]\nonumber \\
 & = & \int d^{2}\alpha\alpha^{\ast k}\alpha^{l}P_{a}^{(s)}(\alpha,t),\label{eq:ordered-moments-def}\end{eqnarray}
where $\{\hat{b}^{\dagger k}\hat{b}^{l}\}_{s}$ denotes the $s$-ordered
product of operators and $a=1,2$ specifies the electronic level.
As a result of the Fourier transform, these moments are obtained equivalently
as derivatives of the corresponding characteristics function, \begin{equation}
\left\langle \left\{ \hat{b}^{\dagger k}(t)\hat{b}^{l}(t)\right\} _{s}\right\rangle _{a}=(-1)^{l}\left[\partial_{\alpha}^{k}\partial_{\alpha^{\ast}}^{l}\underline{P}_{a}^{(s)}(\alpha,t)\right]_{\alpha=0}.\label{eq:moments}\end{equation}

Inserting the mapping of characteristic functions (\ref{eq:map2})
into (\ref{eq:moments}) and applying the Leibniz formula yields the
mapping of initial quantum-statistical moments of the ion's vibration
on the final moments ($t_{0}=0$, and $t\to\infty$): \begin{widetext}
\begin{equation}
\left\langle \left\{ \hat{b}^{\dagger k}(\infty)\hat{b}^{l}(\infty)\right\} _{s}\right\rangle _{2}=P_{2}(0)\left\langle \left\{ \hat{b}^{\dagger k}(0)\hat{b}^{l}(0)\right\} _{s}\right\rangle _{2}+P_{1}(0)\sum_{n=0}^{k}\sum_{m=0}^{l}{k \choose n}{l \choose m}\left\langle \alpha^{\ast(k-n)}\alpha^{l-m}\right\rangle _{p}\left\langle \left\{ \hat{b}^{\dagger n}(0)\hat{b}^{m}(0)\right\} _{s}\right\rangle _{1}.\label{eq:map3}\end{equation}
\end{widetext}The moments of the density of recoil shifts, $\langle\alpha^{\ast k}\alpha^{l}\rangle_{p}$,
act here as weight factors for contributions to the sum of initial
vibrational moments on the rhs. 

Setting $k=0$ and $l=1$ in Eq. (\ref{eq:map3}), the mapping of
the coherent vibrational amplitude is obtained as \begin{equation}
\langle\hat{b}(\infty)\rangle_{2}=\langle\hat{b}(0)\rangle,\label{eq:moment-a}\end{equation}
where the complete quantum-statistical average on the rhs is defined
as\begin{equation}
\langle\ldots\rangle=\sum_{a=1,2}P_{a}\langle\ldots\rangle_{a}={\rm Tr}\left[\hat{\varrho}\ldots\right].\label{eq:trace-def}\end{equation}
Thus the coherent vibration is unaffected by the pump process, as
is the expectation value of the phase-dependent quadrature operator

\begin{equation}
\hat{q}_{\phi}=\frac{1}{\sqrt{2}}\left(\hat{b}\, e^{i\phi}+\hat{b}^{\dagger}e^{-i\phi}\right),\label{eq:def-q}\end{equation}
which reveals the mapping\begin{equation}
\langle\hat{q}_{\phi}(\infty)\rangle_{2}=\langle\hat{q}_{\phi}(0)\rangle.\label{eq:q-map}\end{equation}
Although optical pumping is a highly incoherent process, the initial
coherent amplitude of the ion's oscillation in the trapping potential
is perfectly preserved.

The mean vibrational excitation, however, is altered. Using Eq. (\ref{eq:map3})
with values $k=l=1$ and specifying normally ordered operator products,
i.e. setting $s=-1$, the mapping of the mean vibrational excitation
is obtained as \begin{equation}
\left\langle \hat{n}(\infty)\right\rangle _{2}=\left\langle \hat{n}(0)\right\rangle +P_{1}(0)\bar{n}_{p},\label{eq:moment-a*a}\end{equation}
where $\hat{n}=\hat{b}^{\dagger}\hat{b}$ is the number of vibrational
quanta and the mean number of vibrational quanta added by the pump
process, $\bar{n}_{p}$, is given by Eq. (\ref{eq:n-avg2b}). Thus,
as seen from Eq. (\ref{eq:n-avg2b}), the addition of mean vibrational
excitation by the optical pump process does entirely depend on the
Lamb-Dicke parameters and on the mean number of spontaneously emitted
photons, that is determined by the branching ratio, $\langle\hat{n}_{{\rm ph}}\rangle=\lambda_{1}/\lambda_{2}+1$.
It does \emph{not} depend on the speed of optical pumping that is
determined by the parameters of the pump laser.

Let us now turn to the rms spreads of these properties. Whereas the
expectation value of the phase-dependent quadrature is preserved during
the optical pumping, see Eq. (\ref{eq:q-map}), the corresponding
rms spread is not. The square of this rms spread, i.e. the variance,
is constructed as\begin{equation}
\langle[\Delta\hat{q}_{\phi}]^{2}\rangle=\langle\hat{n}\rangle+\frac{1}{2}+\Re\left[\langle\hat{b}^{2}\rangle e^{2i\phi}\right]-\langle\hat{q}_{\phi}\rangle^{2},\label{eq:qq-map2}\end{equation}
where only the last two terms are phase dependent. From Eq. (\ref{eq:map3})
the mapping of the moment $\langle\hat{b}^{2}\rangle$ is obtained
with the choice $k=0$, $l=2$ as\begin{equation}
\left\langle \hat{b}^{2}(\infty)\right\rangle _{2}=\left\langle \hat{b}^{2}(0)\right\rangle +P_{1}(0)\left\langle \alpha^{2}\right\rangle _{p}.\label{eq:map-b2-def}\end{equation}
Thus, with the mappings (\ref{eq:q-map}) and (\ref{eq:moment-a*a}),
and after completion of pumping, the variance of the quadrature in
the pumped electronic state becomes\begin{equation}
\langle[\Delta\hat{q}_{\phi}(\infty)]^{2}\rangle_{2}=\langle[\Delta\hat{q}_{\phi}(0)]^{2}\rangle+P_{1}(0)\delta q_{p,\phi}^{2},\label{eq:dq-map-def}\end{equation}
where the additional noise $\delta q_{p,\phi}^{2}$ is given in Eq.
(\ref{eq:qq-map}). In accordance with the moments of the density
of recoil shifts, minimum noise is added to the rms spread of the
quadrature at the phase $\phi_{-}=-\phi_{A}/2$, and maximum noise
is added at phase $\phi_{+}=\phi_{-}+\pi/2$.

Finally most relevant is how broad grows the final vibrational number
distribution. The variance of the vibrational quantum number is obtained
from normally ordered moments as\begin{equation}
\langle[\Delta\hat{n}]^{2}\rangle=\langle\hat{b}^{\dagger2}\hat{b}^{2}\rangle-\langle\hat{n}\rangle\left(\langle\hat{n}\rangle-1\right).\label{eq:n-var}\end{equation}
From Eq. (\ref{eq:map3}), using $k=l=2$ and $s=-1$, the mapping
of the required normally-ordered moment is obtained as \begin{widetext}\begin{equation}
\left\langle \hat{b}^{\dagger2}(\infty)\hat{b}^{2}(\infty)\right\rangle _{2}=\left\langle \hat{b}^{\dagger2}(0)\hat{b}^{2}(0)\right\rangle +P_{1}(0)\left\{ 2\Re\left[\left\langle \alpha^{2}\right\rangle _{p}\left\langle \hat{b}^{\dagger2}(0)\right\rangle _{1}\right]+4\left\langle \alpha^{\ast}\alpha\right\rangle _{p}\left\langle \hat{n}(0)\right\rangle _{1}+\left\langle \alpha^{\ast2}\alpha^{2}\right\rangle _{p}\right\} .\label{eq:map3b}\end{equation}
\end{widetext}Thus using Eqs (\ref{eq:moment-a*a}) and (\ref{eq:map3b})
the mapping of the variance of the vibrational quantum number, Eq.
(\ref{eq:n-var}), results as \begin{eqnarray}
\langle[\Delta\hat{n}(\infty)]^{2}\rangle_{2} & = & \langle[\Delta\hat{n}(0)]^{2}\rangle\label{eq:delta-n-map}\\
 & + & P_{1}(0)\left\{ \delta n_{p}^{2}+2\bar{n}_{p}\left[m_{1}+P_{2}(0)m_{2}\right]\right\} .\nonumber \end{eqnarray}
The numbers $m_{1}$ and $m_{2}$ depend on the initial state,\begin{eqnarray}
m_{1} & = & \left[\langle\hat{n}(0)\rangle_{1}+\frac{1}{2}\right]-A\Re\left[\langle\hat{b}^{2}(0)\rangle_{1}e^{-i\phi_{A}}\right],\label{eq:def-m1}\\
m_{2} & = & \frac{\bar{n}_{p}}{2}+\langle\hat{n}(0)\rangle_{1}-\langle\hat{n}(0)\rangle_{2},\label{eq:def-m2}\end{eqnarray}
and $\delta n_{p}^{2}$ is given by Eq. (\ref{eq:np-var2}). 

For example, an atom being initially laser-cooled to its trap ground
state {[}$\langle\hat{n}(0)\rangle_{1,2}=\langle\hat{n}(0)\rangle=0$,
$\langle\hat{b}^{2}(0)\rangle=0$], and also in its electronic ground
state {[}$P_{1}(0)=1$, $P_{2}(0)=0$], has $m_{1}=1/2$ and $m_{2}=\bar{n}_{p}/2$,
the latter not entering the final variance. The mapping of the variance
of the vibrational excitation then simplifies to \begin{equation}
\langle[\Delta\hat{n}(\infty)]^{2}\rangle_{2}=\bar{n}_{p}+\delta n_{p}^{2},\label{eq:delta-n-|0>}\end{equation}
i.e., the variance of the final vibrational distribution equals the
mean plus variance of the recoil displacements.

\section{Results\label{sec:5}}

\subsection{Fluorescence scattering}

As noted before the case of fluorescence scattering is covered by
the model in the limit $\lambda_{2}\to0$, when the trail of spontaneously
emitted photons becomes infinite so that the directions of recoils
in vibrational phase space are efficiently averaged to result in an
isotropic recoil density. However, the resulting recoil density would
correspond to the limit $t\to\infty$, when the diffusion of fluorescence
scattering would have infinitely broadened the density distribution.
Therefore, a direct comparison is impossible, and we restrict ourselves
to a discussion of the asymptotic behavior for small $\lambda_{2}$.

In the limit $\lambda_{2}\to0$ the anisotropy parameter becomes\begin{equation}
A\to\left\{ \begin{array}{cc}
1, & \mbox{if }\underline{w}=1\;\mbox{and}\;\phi_{\underline{w}}=0,\\
0, & \mbox{else}.\end{array}\right.\label{eq:A-limit}\end{equation}
Apart from the exceptional case $\underline{w}(2\nu)=1$ %
\footnote{This case is impossible to realize, at least for resonant pumping
($\Delta=0$), as can be seen from the results of App. \ref{sec:waiting-time-distribution}.%
}, where the waiting times are perfectly synchronized to half the trap
period so that a highly directional scattering occurs ($A\to1$),
the recoil density is isotropic with respect to the variance of its
phase-dependent quadrature ($A\to0$). In more detail, for small $\lambda_{2}$
the anisotropy's asymptotic behavior is given by $A\asymp a\lambda_{2}$
and $\phi_{A}\asymp{\rm const}$, where the coefficient $a$ is \begin{equation}
a=\frac{\underline{w}}{\sqrt{(1-\underline{w})^{2}+4\underline{w}\sin^{2}(\phi_{\underline{w}}/2)}}.\label{eq:a-chico-def}\end{equation}

The asymptotic form of the rms spread of the quadrature becomes then
phase independent, cf. Eq. (\ref{eq:qq-map}), and results as\begin{equation}
\delta q_{p,\phi}\asymp\sqrt{\bar{n}_{p}}.\label{eq:dq-phase-indep}\end{equation}
Like the mean vibrational excitation (\ref{eq:n-avg2b}), the rms
spread of the quadrature is now independent of the laser parameters.
As the mean vibrational excitation asymptotically behaves as\begin{equation}
\bar{n}_{p}\asymp\frac{\eta_{1}^{2}}{\lambda_{2}}\langle(\cos\theta)^{2}\rangle_{1},\label{eq:n-asymp-def}\end{equation}
the rms spread of the quadrature (\ref{eq:dq-phase-indep}) asymptotically
takes on the form,\begin{equation}
\delta q_{p,\phi}\asymp\eta_{1}\sqrt{\langle(\cos\theta)^{2}\rangle_{1}/\lambda_{2}}.\label{eq:dq-asymp-def}\end{equation}
Correspondingly, the rms spread is asymptotically $\delta n_{p}\asymp\bar{n}_{p}$
and also turns independent of the laser parameters. Thus, we expect
an isotropic recoil density in phase space that just weakly depends
on the laser parameters through higher-order moments.

\begin{figure}
\noindent \begin{centering}\hspace*{-22mm}\includegraphics[clip,width=0.68\textwidth,keepaspectratio]{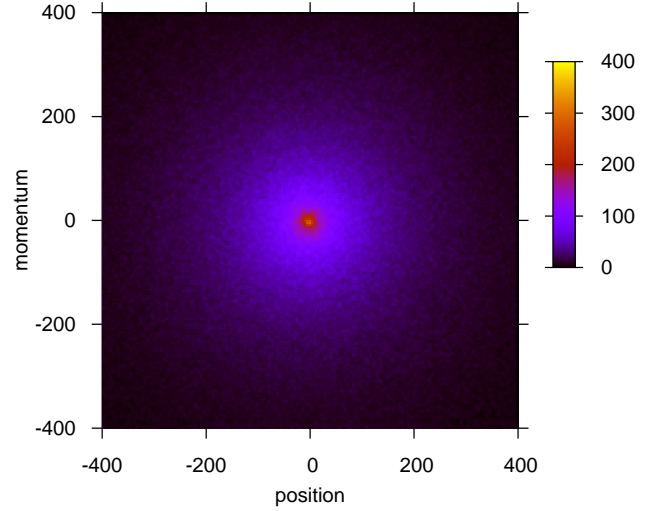}\par\end{centering}

\caption{\label{fig:Recoil-densities-for}Recoil density $p(\alpha)$ in dependence
on position and momentum scaled as $x/(\eta_{1}\Delta x_{0})$ and
$p/(\eta_{1}\Delta p_{0})$, respectively, with trap ground-state
uncertainties being $\Delta x_{0}\Delta p_{0}=\hbar/2$. Parameters
are $\lambda_{2}=10^{-5}$, $\eta_{2}/\eta_{1}=0.75$, $\tilde{\nu}=0.16$,
$\Delta=0$, and $S=25$ (1.000.000 phase-space shifts sampled on
a 200$\times$200 grid). Transition dipole moments are assumed to
be perpendicular to the chosen motional axis. }
\end{figure}

\begin{figure}
\includegraphics[width=0.48\textwidth]{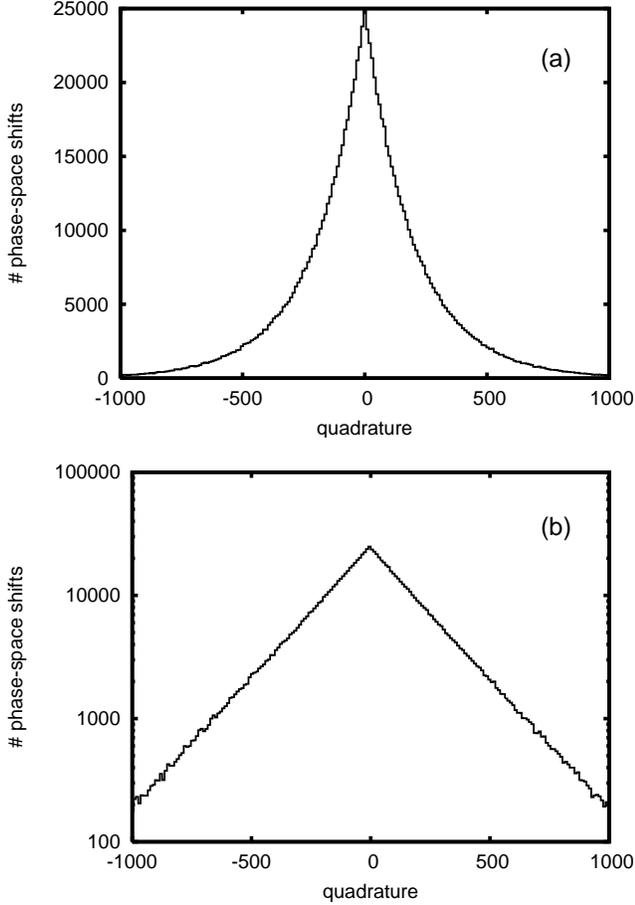}

\caption{\label{fig:Quadrature-distributions-along}(a) Phase-independent
distributions of quadratures of $p(\alpha)$ for (resonant) fluorescence
scattering; parameters are the same as in Fig. \ref{fig:Recoil-densities-for}
(1.000.000 phase-space shifts sampled in 200 bins in the interval
{[}-1000:1000]); (b) Logarithmic plot of the same function.}
\end{figure}

Numerical evaluation of Eq. (\ref{eq:displ-distr}) for $t\to\infty$
($t_{0}=0$) and $\lambda_{2}=10^{-5}$ in resonant excitation ($\Delta=0$)
reveals the isotropic recoil density in phase space as shown in Fig.
\ref{fig:Recoil-densities-for}, which is independent of the preset
laser saturation $S=(|\kappa|/\gamma)^{2}$, in agreement with previous
considerations. Given the infinite trail of spontaneously emitted
photons, each photon contributing to a random recoil, one might presume
the validity of the central-limit theorem that would predict a perfectly
Gaussian profile of the recoil density. However, the individual recoil
shifts in phase space are not precisely independent random variables.
The phases, i.e. directions in phase space, of subsequent recoils
are correlated by means of the waiting-time distribution (\ref{eq:w-def}).
The detailed form of this distribution seems to be lost in the averaging
process, its main characteristics, however, a finite mean waiting
time between subsequent photon emissions, leaves a correlation between
subsequent recoil shifts. This effect appears as substantial deviation
from a Gaussian profile in the distribution of quadratures shown in
Fig. \ref{fig:Quadrature-distributions-along} (a). In fact, this
recoil density perfectly approaches a $\exp(-|x|)$ function as can
be seen from the corresponding logarithmic plot in Fig. \ref{fig:Quadrature-distributions-along}
(b).

It has been shown that fluorescence scattering of a free atom generates,
after many photon emissions, a Gaussian momentum distribution \cite{pusep,mandel}.
The above results show that this situation corresponds to the limit
$\nu\to0$, i.e., when switching off the trap potential. Then, the
spectral waiting-time distribution enters Eqs. (\ref{eq:A-modulus})
and (\ref{eq:A-phase}) at zero frequency and equals unity due to
the normalization of the waiting-time distribution, $\underline{w}(0)=1$.
Thus, in this particular case the laser parameters do not appear,
and neither does the waiting-time distribution -- except its normalization
property. Each photon recoil shifts the atomic momentum in the momentum's
original direction, and the length of each shift is a statistically
independent random number, whose statistics is given by the dipole
radiation characteristics. The application of the central-limit theorem
is here perfectly justified given a sufficiently large number of spontaneous
photon emissions.

\subsection{Optical pumping with few photon emissions}

On the other hand, in the opposite limit $\lambda_{2}\to1$, the atom
spontaneously emits one photon only and reaches its final pumped state.
In this limit, the joint photon emission probability density (\ref{eq:photon-joint-prob})
becomes\begin{equation}
\lim_{\lambda_{2}\to1}w_{n}(t_{n},\ldots,t_{1})=\delta_{n,1}w(t_{1}-t_{0}),\label{eq:w-lim-only-one}\end{equation}
from which the recoil density (\ref{eq:displ-distr}) is derived for
completed pumping ($t\to\infty$, $t_{0}=0$) and making use of (\ref{eq:disp-traj})
as\begin{equation}
p(\alpha)=\int_{0}^{\infty}dt\int ds\mu_{2}(s)w(t)\delta[\alpha-i\eta_{2}s\exp(i\nu t)].\label{eq:density-anisotropic}\end{equation}

For high enough laser saturation ($S>1$) the waiting-time distribution
decays at the rate on the order of $\gamma$. In an experiment, the
trap frequency is typically much lower, $\nu\ll\gamma$. Thus, to
very good approximation we may discard in Eq. (\ref{eq:density-anisotropic})
the comparably slow time dependence of the direction of the recoil
shift, $e^{i\nu t}\approx1$, perform the integral over the two-dimensional
delta function and obtain \begin{equation}
p(\alpha)\approx\delta(q)\mu_{2}\left(\frac{p}{\eta_{2}}\right)/\eta_{2},\label{eq:density-anisotropic2}\end{equation}
where $\alpha=q+ip$. Thus the recoil density extends only in direction
of momentum in phase space, reproducing the profile of the dipole
radiation characteristics. This density distribution agrees with the
profile associated with a free atom after single-photon scattering
\cite{mandel}.

\begin{figure}
\hspace*{-22mm}\includegraphics[width=0.68\textwidth,keepaspectratio]{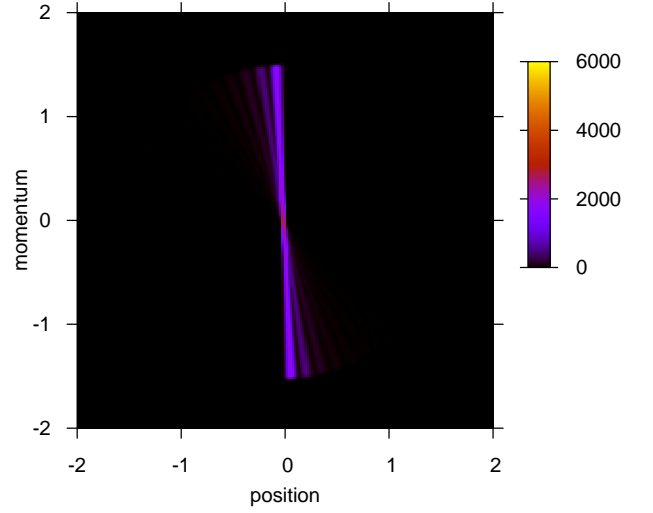}

\caption{\label{fig:Optical-pumping-with}Recoil density $p(\alpha)$ versus
position and momentum scaled as $x/(\eta_{1}\Delta x_{0})$ and $p/(\eta_{1}\Delta p_{0})$,
respectively, with trap ground-state uncertainties being $\Delta x_{0}\Delta p_{0}=\hbar/2$.
Parameters are $\lambda_{1}=10^{-5}$, $\eta_{2}/\eta_{1}=0.75$,
$\tilde{\nu}=0.16$, $\Delta=0$, and $S=25$ (1.000.000 phase-space
shifts sampled on a 200$\times$200 grid). Transition dipole moments
are assumed to be perpendicular to the selected motional axis.}
\end{figure}

\begin{figure}
\begin{centering}\includegraphics[width=0.48\textwidth,keepaspectratio]{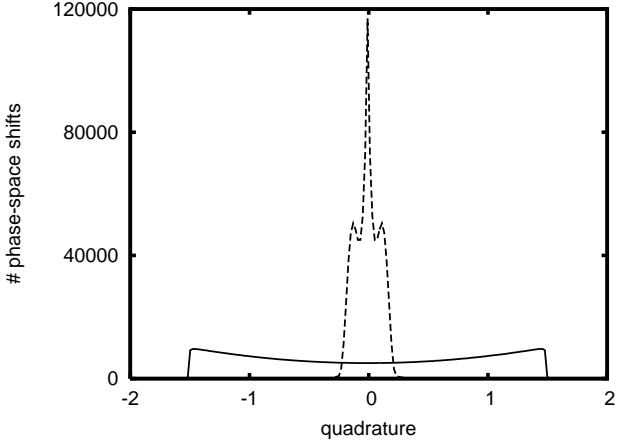}\par\end{centering}

\caption{\label{fig:Quadrature-distributions-for}Distributions of quadratures
of $p(\alpha)$ versus the scaled quadrature corresponding to maximum
(solid) and minimum (dashed) rms quadrature fluctuations. Parameters
are those of Fig. \ref{fig:Optical-pumping-with} (1.000.000 phase-space
shifts sampled in 200 bins on the interval $[-2,2]$).}
\end{figure}

In Fig. \ref{fig:Optical-pumping-with} we approach this limiting
case for the branching ratio $\lambda_{1}=10^{-5}$. The almost unidirectional
scattering in phase space is well observed. Moreover, for the corresponding
quadrature, the distribution of maximum rms fluctuation shows a dipole
radiation characteristics, see solid curve in Fig. \ref{fig:Quadrature-distributions-for}.
The secondary and tertiary lobes in the recoil density of Fig. \ref{fig:Optical-pumping-with}
mark delayed spontaneous emission, modulated by the laser-driven damped
Rabi cycles on the $|1\rangle\leftrightarrow|3\rangle$ transition.

\subsection{Laser-saturation dependence of the anisotropy}

The anisotropy can be written as a function of three variables,\begin{equation}
A(\lambda_{1},\underline{w},\phi_{\underline{w}})=\frac{(1-\lambda_{1})\underline{w}}{\sqrt{(1-\lambda_{1}\underline{w})^{2}+4\lambda_{1}\underline{w}\sin^{2}(\phi_{\underline{w}}/2)}}.\label{eq:A-def-repeat}\end{equation}
Two of these variables, $\underline{w}$ and $\phi_{\underline{w}}$,
depend on the laser parameters, i.e., on detuning and saturation of
transition $|1\rangle\leftrightarrow|3\rangle$. For resonant pumping
($\Delta=0$), these functions of laser saturation $S$ are obtained
as {[}see Eqs (\ref{eq:w-reson}) and (\ref{eq:phiw-reson})]\begin{eqnarray}
\underline{w}(S) & = & \frac{S}{\sqrt{\left(1+\tilde{\nu}^{2}\right)\left[\left(S-\tilde{\nu}^{2}\right)^{2}+4\tilde{\nu}^{2}\right]}},\label{eq:w(S)}\\
\phi_{\underline{w}}(S) & = & \arctan\left[\frac{\tilde{\nu}\left(S+2-\tilde{\nu}^{2}\right)}{S-3\tilde{\nu}^{2}}\right],\label{eq:tanphi(S)}\end{eqnarray}
with $\tilde{\nu}=2\nu/\gamma$. 

Thus, the anisotropy $A(S)$ depends on laser saturation with the
branching ratio $\lambda_{2}$ and the scaled trap frequency $\tilde{\nu}$
being parameters. With Eqs. (\ref{eq:w(S)}) and (\ref{eq:tanphi(S)})
inserted into (\ref{eq:A-def-repeat}), this function is \begin{equation}
A(S)=\frac{\lambda_{2}S}{\sqrt{\left(\lambda_{2}S-3\tilde{\nu}^{2}\right)^{2}+\tilde{\nu}^{2}\left(S+2-\tilde{\nu}^{2}\right)^{2}}}.\label{eq:A(S)-def-first}\end{equation}
Its saturated value is\begin{equation}
\lim_{S\to\infty}A(S)=\frac{\lambda_{2}}{\sqrt{\lambda_{2}^{2}+\tilde{\nu}^{2}}},\label{eq:A(S)-sat-value}\end{equation}
which vanishes for resonant fluorescence scattering, $\lambda_{2}\to0$.
Likewise, the phase may be written as the saturation-dependent function\begin{equation}
\tan\phi_{A}(S)=\frac{\sin\phi_{w}(S)}{\cos\phi_{w}(S)-\lambda_{1}\underline{w}(S)}.\label{eq:tanphi(S)-def-first}\end{equation}

\begin{figure}
\begin{centering}\includegraphics[width=0.42\textwidth]{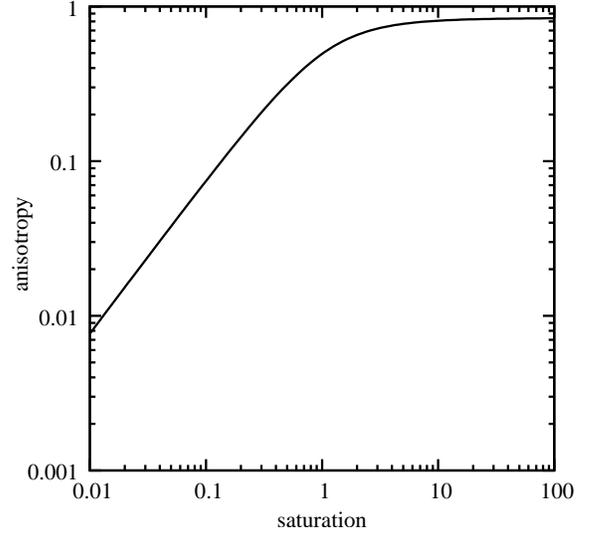}\par\end{centering}

\caption{\label{fig:anisotropy}Anisotropy $A$ as a function of the laser
saturation $S$ for resonant pumping ($\Delta=0$); parameters are
$\tilde{\nu}=0.16$, $\lambda_{1}=0.75$. }
\end{figure}

The anisotropy, an example of which is shown in Fig. \ref{fig:anisotropy},
may have a local maximum in form of a typical under-damped peak. However,
it can be shown that this maximum appears at the saturation value
\begin{equation}
S_{{\rm max}}=\frac{\tilde{\nu}^{4}+5\tilde{\nu}^{2}+4}{\tilde{\nu}^{2}+3\lambda_{2}-2}\label{eq:S-peak-def}\end{equation}
only if the condition \begin{equation}
\lambda_{2}>\frac{2-\tilde{\nu}^{2}}{3}\label{eq:lambda-cond}\end{equation}
holds. Otherwise the global maximum is attained at full saturation
$S\to\infty$. For example, given the typical value $\tilde{\nu}=2\nu/\gamma=0.16$,
a branching ratio $\lambda_{2}\gtrsim0.66$ is required to observe
the maximum anisotropy at extremely high laser saturation. Typically,
this regime is inaccessible in experiment, so that usually the anisotropy
is a monotonically increasing function of the saturation, as shown
in Fig. \ref{fig:anisotropy}.

\subsection{Optimized optical pumping}

In many applications one intends to minimize the disturbance of optical
pumping on an initially given vibrational quantum state, in order
to only affect the electronic system in the desired way. Optimization
depends, however, on the specific type of disturbance that is supposed
to be minimized. As we have seen in Sec. \ref{sec:3.1}, the various
vibrational moments depend quite differently on the laser parameters,
some even being independent of the kind of laser excitation. Thus,
we may present optimum values for optical pumping in a common case.
To discuss the worst-case scenario, we start from an atom with initial
population in the unpumped state $P_{1}(0)=1$.

For keeping the rise of the mean vibrational excitation (i.e. heating)
as small as possible during optical pumping, as indicated by Eq. (\ref{eq:moment-a*a}),
the moment $\bar{n}_{p}$, i.e., the mean recoil-induced excitation,
has to be minimized. This moment does not depend on laser parameters,
cf. Eq. (\ref{eq:n-avg2b}), but is a fixed value determined by the
Lamb--Dicke parameters, the branching ratio and the second moment
of the dipole radiation characteristics. It is thus irrelevant whether
optical pumping is performed resonantly or off-resonantly, or whether
a low or high saturation is employed. 

The width of the resulting distribution over vibrational quantum numbers,
however, does depend on the laser parameters. While the atom is initially
in its vibrational ground state $|n=0\rangle$, the final spread of
vibrational quantum numbers in the pumped state is given by Eq. (\ref{eq:delta-n-|0>})
as\begin{eqnarray}
 &  & \langle[\Delta\hat{n}(\infty)]^{2}\rangle_{2}=\bar{n}_{p}^{2}+\frac{\lambda_{1}}{\lambda_{2}^{2}}\eta_{1}^{4}\langle(\cos\theta)^{2}\rangle_{1}^{2}\label{eq:dn-optimized-def}\\
 &  & \quad+\left(\eta_{2}^{4}C_{2}+\frac{\lambda_{1}}{\lambda_{2}}\eta_{1}^{4}C_{1}\right)+2\bar{n}_{p}\frac{\lambda_{1}}{\lambda_{2}}\eta_{1}^{2}\langle(\cos\theta)^{2}\rangle_{1}A\cos\phi_{A}\nonumber \end{eqnarray}
where we have defined\begin{equation}
C_{1,2}=\langle(\cos\theta)^{4}\rangle_{1,2}-\langle(\cos\theta)^{2}\rangle_{1,2}^{2}.\label{eq:def-strange-coeg}\end{equation}
This is minimized for $A\cos\phi_{A}$ being as negative as possible.
For resonant pumping ($\Delta=0$) the anisotropy parameters are only
functions of the saturation $S$. Thus, we may search for the minimum
of $A(S)\cos\phi_{A}(S)$. Extrema of this function follow from the
condition\begin{equation}
\frac{A^{\prime}(S)}{A(S)}=\phi_{A}^{\prime}(S)\tan\phi_{A}(S),\label{eq:funny-condition}\end{equation}
where the primes indicate derivatives. The equation can be shown to
have only negative solutions, whereas the saturation by definition
is a semi-positive number. Since the global minimum is located at
$S=0$, where $A\cos\phi_{A}=0$, a minimum spread of the vibrational
numbers is obtained when saturation is kept as low as possible. Thus,
slow optical pumping is required for minimum width of the vibrational
number distribution. Optimization with respect to other moments, however,
requires other values of saturation.

\section{Motivation and Context\label{sec:Motivation-and-Context}}

\subsection{The center-of-mass effects in optical pumping}

Optical pumping is a time-honoured subject. A wealth of experiments
have required various approaches of modelling. However, these theories
usually limited their models to the description of the atom's electronic
dynamics driven by the light. The center-of-mass dynamics, modified
by the unavoidable radiative recoil, was usually neglected on the
basis of the assumption that atomic collisions and in particular the
spontaneous emission intrinsically tied to the pumping process would
average an atom's state to an unspectacular slightly broadened Gaussian
distribution in phase space that might at most render energy levels
and transitions broadened by an almost unobservable amount. 

The results of the above calculations show that this attitude is an
oversimplification. For complete description of OP, the demonstrated
modelling of the recoil effects is indispensable. These effects leave,
in general, a pumped atom in a non-Gaussian state with an anisotropic
distribution in phase space. The model reveals a continuous transit
from maximum asymmetry to full symmetry going along with the number
of reiterated scattering events increasing up to the final pumping
event. Consequently, the center-of-mass effect of optical pumping
is linked in a natural way to that of resonance scattering on the
pump-excited line: The latter phenomenon is described by an infinite
series of scattering events, lacking the final spontaneous decay into
the pumped state. When the pump light is detuned off resonance, the
accumulated recoil turns out to be a detrimental heating contribution
to the effect of laser sideband cooling of the atom weakly bound in
a trap. (The wanted contribution is from the recoil by the pump light.)
For an atom strongly bound to the trap, however, the momentum of recoil
is tranferred to the entire trap. This process is analogous with the
origin of narrow-band fluorescence emission in the Mössbauer effect.

\subsection{Optical pumping of vapors}

Optical pumping of trapped or free atomic vapors can be described
by our model, except for ultra low temperature when the gas is degenerate.
Elastic collisions will then generate scattering among vibrational
levels, to effect thermalisation of the vapor.

The atoms will be in a non-equilibrium state after optical pumping.
As with the electronic inversion, this also applies to the vibrational
state. Both the particular shape and the anisotropy of the density
of recoil shifts is transmitted in optical pumping to the final vibrational
phase-space distribution. Such a distribution, in general, is far
from a thermal distribution corresponding to equilibrium. In particular,
the anisotropy in phase space indicates that kinetic and potential
energies of the ensemble are unbalanced, which manifests itself by
the atomic cloud breathing at twice the trap frequency.

Atomic collisions redistribute populations in the vibrational levels
and thus rebalance kinetic versus potential energy to arrive at a
thermalized equilibrium state. This process happens on a time scale
that is determined by the atomic scattering cross section. Given that
this time scale is large enough to be detected experimentally, the
observation of the cloud should reveal damping of the cloud's breathing.

\subsection{Observability of anisotropy and non-Gaussian shape in phase space}

The prominent features of the results presented here are the anisotropy
and the non-Gaussian shape of the density of recoil shifts in phase
space. They can be directly observed by using a simplified version
of a method, proposed for the reconstruction of the vibrational quantum
state \cite{wavo,wavo2}, that has been implemented experimentally
in recent years \cite{monroe}.

After optical pumping a ground-state atom whose initial phase-space
distribution is rotationally symmetric (e.g. any vibrational number
state or incoherent mixtures of these states), the $|1\rangle\leftrightarrow|2\rangle$
transition is bichromatically driven in stimulated Raman configuration
on the first red and the first blue vibrational sideband. Application
of this laser light for a finite and adjustable time duration $\tau$
maps information on the distribution of quadratures onto the occupation
of the ground state $|1\rangle$. Given a laser phase shift $\phi$
introduced in one of the two Raman transitions, the ground state occupation
reads \cite{wavo,wavo2}\begin{equation}
P_{1}(\tau,\phi)=\frac{1}{2}-\frac{1}{2}\int dqe^{-iq\tau}p(q,\phi),\label{eq:tomo-get-p-def}\end{equation}
where the distribution $p(q,\phi)$ is the probability density of
observing the quadrature $q$ at phase $\phi$, as shown in Figs \ref{fig:Quadrature-distributions-along}
and \ref{fig:Quadrature-distributions-for} in Sec. \ref{sec:5}.

The occupation in the ground state $P_{1}$ is obtained by probing
the atom for fluorescence of the transition from $|1\rangle$ to an
auxiliary level. Thus, varying the time duration $\tau$ and Fourier
transforming the measured time-dependent ground state occupation,
all distributions of the phase-dependent quadrature are obtained.
The appearance of phase-dependent distributions of quadratures would
prove the predicted anisotropy of the density of recoil shifts. In
addition, the shape of the measured distributions could be tested
for the deviation from a Gaussian profile.

\section{Summary and conclusions\label{sec:6}}

In this paper it was shown how recoil effects that naturally appear
with optical pumping are analytically modelled for a harmonically
bound atom. Due to the trail of spontaneously emitted photons being
finite, a quantum trajectory-type of approach is well adapted to this
analytic procedure. All effects on the quantum motion of the atom
depend on a single function, the recoil density in phase space. Its
moments are obtained analytically from the dipole radiation characteristics
and the waiting-time distribution of the atom's light scattering. 

It was shown that even approaching the case of fluorescence scattering
the recoil density does not become a Gaussian function in the motional
phase space, as one would expect when taking resort to the central-limit
theorem. Instead, it is a distribution of the form $\exp(-|x|)$,
that emerges from correlations of emission times and the concomitant
correlations of the recoil directions in phase space of subsequent
spontaneous photon emissions. In the opposite extreme, when only few
photons are scattered during the full pump cycle, the recoil density
is characterized by the spatial distribution of dipole radiation.

In general, the recoil density is not isotropically distributed but
reveals a pronounced directional structure. For resonant pumping,
this anisotropy can be maximized adjusting the laser saturation --
in most relevant cases by increasing it.

Optimum laser saturation and detuning for minimum detrimental heating
depend on the particular property that is to be protected from the
recoil effects. Whereas the mean increase of vibrational excitation
cannot be varied, the rms fluctuations of the vibrational number can
be affected by the laser parameters, as is the case also for the rms
fluctuations of the phase-dependent quadrature. However, optimum laser
parameters depend on the initial motional quantum state and on the
particular phase of a quadrature to become protected. There exists
a complex interplay among the initial motional quantum state, the
particular fluctuations to be minimized, and the laser parameters
that fit in with the given requirements.

\begin{acknowledgments}
S.W. and P.E.T. acknowledge support by FONDECYT grant no. 7060187,
S.W. acknowledges support by FONDECYT grant no. 1051042.
\end{acknowledgments}
\appendix

\section{Derivatives of the characteristic function for instantaneous recoil
shifts\label{app:deriv-charac}}

The characteristic function associated with the probability density
of instantaneous recoil shifts is

\begin{eqnarray*}
\underline{p}_{a}(\alpha,t) & = & \underline{\mu}_{a}\left[2\eta_{a}|\alpha|\cos(\nu t-\arg\alpha)\right]\\
 & = & \underline{\mu}_{a}\left[\eta_{a}\left(\alpha^{\ast}e^{i\nu t}+\alpha e^{-i\nu t}\right)\right],\end{eqnarray*}
where $\nu$ is the vibrational frequency. Its $(k,l)$-fold derivatives
are

\begin{eqnarray*}
 &  & \left.\left(\frac{\partial}{\partial\alpha^{\ast}}\right)^{l}\left(\frac{\partial}{\partial\alpha}\right)^{k}\underline{p}_{a}(\alpha,t)\right|_{\alpha=0}\\
 &  & =\left.\left(\frac{\partial}{\partial\alpha^{\ast}}\right)^{l}\left(\eta_{a}e^{-i\nu t}\right)^{k}\underline{\mu}_{a}^{(k)}\left[\eta_{a}\left(\alpha^{\ast}e^{i\nu t}+\alpha e^{-i\nu t}\right)\right]\right|_{\alpha=0}\\
 &  & =\left(\eta_{a}e^{i\nu t}\right)^{l}\left(\eta_{a}e^{-i\nu t}\right)^{k}\underline{\mu}_{a}^{(k+l)}(0),\end{eqnarray*}
where \begin{eqnarray*}
\underline{\mu}_{a}^{(k+l)}(0) & = & (-i)^{k+l}\int_{-1}^{1}ds\mu_{a}(s)s^{k+l}\\
 & = & (-i)^{k+l}\left\langle (\cos\theta)^{k+l}\right\rangle _{a},\end{eqnarray*}
so that \begin{eqnarray}
\left.\left(\frac{\partial}{\partial\alpha^{\ast}}\right)^{l}\left(\frac{\partial}{\partial\alpha}\right)^{k}\underline{p}_{a}(\alpha,t)\right|_{\alpha=0} & = & (-i\eta_{a})^{k+l}e^{i\nu t(l-k)}\nonumber \\
 &  & \times\left\langle (\cos\theta)^{k+l}\right\rangle _{a}.\label{eq:p-derivs-result}\end{eqnarray}

The dipole radiation characteristics is obtained from Eq. (\ref{eq:dipole-charac})
as\begin{equation}
\mu_{a}(s)=\frac{3}{8\pi}\int_{0}^{2\pi}d\phi\left\{ 1-\cos^{2}\alpha\right\} ,\label{eq:dip-charac-app}\end{equation}
where $\alpha$ is the angle between $\vec{n}_{a}$ and $\vec{n}(\Omega)$
and thus \begin{equation}
\cos\alpha=\cos\theta\cos\theta_{a}+\sin\theta\sin\theta_{a}\cos(\phi-\phi_{a}),\label{eq:cosalpha}\end{equation}
where $\theta_{a}$ and $\phi_{a}$ are the spherical angles of the
transient dipole moment $\vec{d}_{3a}=\langle3|\hat{\vec{d}}|a\rangle$.
Inserting Eq. (\ref{eq:cosalpha}) into (\ref{eq:dip-charac-app})
and performing the integration the dipole characteristics results
as\begin{equation}
\mu_{a}(s)=\frac{3}{8}\left[1+\cos^{2}\theta_{a}+s^{2}\left(1-3\cos^{2}\theta_{a}\right)\right].\label{eq:dip-charac-app2}\end{equation}
It is thus a symmetric function, $\mu_{a}(-s)=\mu_{a}(s)$, that depends
on the angle $\theta_{a}$ between the transition dipole moment and
the direction of motion.

\section{Photon-counting statistics\label{app:photon-stat}}

From the joint probability of spontaneous photon emissions, the photon-counting
statistics derives from the probability for $n$ spontaneous emissions
during the time interval $[t,t_{0}]$, \begin{equation}
P_{n}(t,t_{0})=\int_{t_{0}}^{t}dt_{n}\ldots\int_{t_{0}}^{t_{2}}dt_{1}w_{n}(t_{n},\ldots,t_{1}),\quad(n>0)\label{eq:photon-count1}\end{equation}
and $P_{0}(t,t_{0})=1-\sum_{n=1}^{\infty}P_{n}(t,t_{0})$. The photon-counting
statistics for a complete optical-pumping process, $P_{n}$, is obtained
in the limit $P_{n}=\lim_{t\to\infty}P(t,t_{0})$ which reads\begin{eqnarray}
P_{n} & = & \int_{t_{0}}^{\infty}dt_{n}\ldots\int_{t_{0}}^{t_{2}}dt_{1}w_{n}(t_{n},\ldots,t_{1})\nonumber \\
 & = & \lambda_{2}\lambda_{1}^{n-1}\int_{t_{0}}^{\infty}dt_{n}\ldots\int_{t_{0}}^{t_{2}}dt_{1}w\left(t_{n}-t_{n-1}\right)\ldots\nonumber \\
 &  & \qquad\ldots w(t_{1}-t_{0}),\label{eq:photon-count2}\end{eqnarray}
where Eq. (\ref{eq:photon-joint-prob}) has been used. The waiting-time
distributions are defined for positive time arguments only, so that
we set $w(t)=0$ for $t<0$ and extend the upper integration limits
to infinity. With the substitution of integration variables $\tau_{k}=t_{k}-t_{k-1}$
($k=1,\ldots,n$) the photon-count statistics writes\begin{eqnarray*}
P_{n} & = & \lambda_{2}\lambda_{1}^{n-1}\int_{0}^{\infty}d\tau_{n}\ldots\int_{0}^{\infty}d\tau_{1}w(\tau_{n})\ldots w(\tau_{1})\\
 & = & \lambda_{2}\lambda_{1}^{n-1}\quad(n>0),\end{eqnarray*}
and $P_{0}=0$. Here, the unit normalization of the waiting-time distribution
has been used.

The average number of spontaneously emitted photons is therefore\begin{eqnarray}
\langle\hat{n}_{{\rm ph}}\rangle & = & \sum_{n=0}^{\infty}nP_{n}=\sum_{n=1}^{\infty}n\lambda_{2}\lambda_{1}^{n-1},\end{eqnarray}
which can be expressed as the derivative\[
\lambda_{2}\frac{\partial}{\partial\lambda_{1}}\sum_{n=1}^{\infty}\lambda_{1}^{n}=\lambda_{2}\frac{\partial}{\partial\lambda_{1}}\frac{\lambda_{1}}{1-\lambda_{1}}=\frac{1}{\lambda_{2}},\]
so that the average photon number is\begin{equation}
\langle\hat{n}_{{\rm ph}}\rangle=\frac{1}{\lambda_{2}}.\label{eq:n-avg}\end{equation}

The rms spread of photon numbers is obtained via the second-order
moment\begin{eqnarray*}
\langle\hat{n}_{{\rm ph}}^{2}\rangle & = & \sum_{n=1}^{\infty}n^{2}\lambda_{2}\lambda_{1}^{n-1}\\
 & = & \lambda_{2}\frac{\partial}{\partial\lambda_{1}}\sum_{n=1}^{\infty}n\lambda_{1}^{n}\\
 & = & \lambda_{2}\frac{\partial}{\partial\lambda_{1}}\left(\frac{\partial}{\partial\lambda_{1}}\lambda_{1}-1\right)\sum_{n=1}^{\infty}\lambda_{1}^{n}\\
 & = & \lambda_{2}\frac{\partial}{\partial\lambda_{1}}\left(\frac{\partial}{\partial\lambda_{1}}\lambda_{1}-1\right)\frac{\lambda_{1}}{1-\lambda_{1}}\\
 & = & \lambda_{2}\frac{\partial}{\partial\lambda_{1}}\left(\frac{\partial}{\partial\lambda_{1}}\frac{\lambda_{1}^{2}}{1-\lambda_{1}}-\frac{\lambda_{1}}{1-\lambda_{1}}\right)\\
 & = & \lambda_{2}\frac{\partial}{\partial\lambda_{1}}\frac{\lambda_{1}}{(1-\lambda_{1})^{2}}\\
 & = & \frac{1+\lambda_{1}}{\lambda_{2}^{2}}\end{eqnarray*}
Therefore, the rms spread reads\[
\langle\Delta\hat{n}_{{\rm ph}}^{2}\rangle=\frac{\lambda_{1}}{\lambda_{2}^{2}}.\]
Its relative value reads\[
\langle\Delta\hat{n}_{{\rm ph}}^{2}\rangle/\langle\hat{n}_{{\rm ph}}\rangle=\frac{\lambda_{1}}{\lambda_{2}}\]
which indicates sub- or super-Poissonian photon statistics, dependent
on the branching ratios.

\section{Moments of the density of recoil shifts\label{app:a2mom}}

The function to be evaluated is given by

\begin{eqnarray}
\langle\alpha^{\ast k}\alpha^{l}\rangle_{p}=(-1)^{l}\sum_{n=1}^{\infty}\int_{t_{0}}^{\infty}dt_{n}\ldots\int_{t_{0}}^{t_{2}}dt_{1}w_{n}(t_{n},\ldots,t_{1})\nonumber \\
\times\left\{ \partial_{\alpha}^{k}\partial_{\alpha^{\ast}}^{l}\left[\underline{p}_{2}(\alpha,t_{n})\underline{p}_{1}(\alpha,t_{n-1})\ldots\underline{p}_{1}(\alpha,t_{n-1})\right]\right\} _{\alpha=0}.\quad\label{eq:P-moments2}\end{eqnarray}

\subsection{Moment $\langle\alpha^{2l}\rangle_{p}$}

From identity (\ref{eq:P-moments2}), the moment $\langle\alpha^{2l}\rangle_{p}$
is given by the expression\begin{eqnarray}
\langle\alpha^{2l}\rangle_{p} & = & \sum_{n=1}^{\infty}\int_{t_{0}}^{\infty}dt_{n}\ldots\int_{t_{0}}^{t_{2}}dt_{1}w_{n}(t_{n},\ldots,t_{1})\label{eq:a2-mom}\\
 & \times & \left\{ \partial_{\alpha^{\ast}}^{2l}\left[\underline{p}_{2}(\alpha,t_{n})\underline{p}_{1}(\alpha,t_{n-1})\ldots\underline{p}_{1}(\alpha,t_{n-1})\right]\right\} _{\alpha=0}.\nonumber \end{eqnarray}
The non-vanishing second derivative has to be applied to all $n$
factors resulting in $n$ terms again. The first term is\begin{eqnarray*}
 &  & \sum_{n=1}^{\infty}\int_{t_{0}}^{\infty}dt_{n}\ldots\int_{t_{0}}^{t_{2}}dt_{1}w_{n}(t_{n},\ldots,t_{1})\\
 &  & \times\left\{ \left[\left[\partial_{\alpha^{\ast}}^{2l}\underline{p}_{2}(\alpha,t_{n})\right]\underline{p}_{1}(\alpha,t_{n-1})\ldots\underline{p}_{1}(\alpha,t_{n-1})\right]\right\} _{\alpha=0},\end{eqnarray*}
where the derivative is $(-i\eta_{2})^{2l}\langle(\cos\theta)^{2l}\rangle_{2}e^{2il\nu t_{n}}$
, cf. Eq. (\ref{eq:p-derivs-result}), so that it reads\begin{eqnarray}
(-i\eta_{2})^{2l}\langle(\cos\theta)^{2l}\rangle_{2}\sum_{n=1}^{\infty}\int_{t_{0}}^{\infty}dt_{n}\ldots\int_{t_{0}}^{t_{2}}dt_{1}\nonumber \\
\times w_{n}(t_{n},\ldots,t_{1})e^{2il\nu t_{n}}.\label{eq:a2-mom-term1}\end{eqnarray}
The sum of the other $n-1$ terms is\begin{eqnarray}
 &  & \sum_{n=1}^{\infty}\int_{t_{0}}^{\infty}dt_{n}\ldots\int_{t_{0}}^{t_{2}}dt_{1}w_{n}(t_{n},\ldots,t_{1})\sum_{p=1}^{n-1}\left\{ \underline{p}_{2}(\alpha,t_{n})\right.\nonumber \\
 &  & \times\left.\underline{p}_{1}(\alpha,t_{n-1})\ldots\left[\partial_{\alpha^{\ast}}^{2l}\underline{p}_{1}(\alpha,t_{p})\right]\ldots\underline{p}_{1}(\alpha,t_{n-1})\right\} _{\alpha=0}\nonumber \\
 &  & =(-i\eta_{1})^{2l}\langle(\cos\theta)^{2l}\rangle_{1}\sum_{n=1}^{\infty}\sum_{p=1}^{n-1}\int_{t_{0}}^{\infty}dt_{n}\ldots\int_{t_{0}}^{t_{2}}dt_{1}\nonumber \\
 &  & \qquad\times w_{n}(t_{n},\ldots,t_{1})e^{2il\nu t_{p}},\label{eq:a2-mom-otherterms}\end{eqnarray}
where the derivative was identified as $(-i\eta_{1})^{2l}\langle(\cos\theta)^{2l}\rangle_{1}e^{2il\nu t_{p}}$.

In both expressions (\ref{eq:a2-mom-term1}) and (\ref{eq:a2-mom-otherterms})
the $n$-fold integral\[
\int_{t_{0}}^{\infty}dt_{n}\ldots\int_{t_{0}}^{t_{2}}dt_{1}w_{n}(t_{n},\ldots,t_{p},\ldots,t_{1})e^{2il\nu t_{p}}\]
with $p=1,\ldots,n$ appears. Inserting the definition of the joint
probability density for $n$ spontaneous emissions (\ref{eq:photon-joint-prob}),
we find\begin{equation}
\lambda_{2}\lambda_{1}^{n-1}\int_{t_{0}}^{\infty}dt_{n}\ldots\int_{t_{0}}^{t_{2}}dt_{1}w(t_{n}-t_{n-1})\ldots w(t_{1}-t_{0})e^{2il\nu t_{p}}.\label{eq:n-integral2}\end{equation}
In the vein of Eq. (\ref{eq:photon-count2}), the upper limits of
the integrals are set to infinity. With substitution of the $n$ integration
variables, $\tau_{n}=t_{n}-t_{n-1}$, expression (\ref{eq:n-integral2})
gives\begin{eqnarray*}
 &  & \lambda_{2}\lambda_{1}^{n-1}e^{2il\nu t_{0}}\int_{0}^{\infty}d\tau_{n}\ldots\int_{0}^{\infty}d\tau_{1}w(\tau_{n})\ldots w(\tau_{1})\\
 &  & \qquad\times e^{2il\nu(\tau_{p}+\ldots+\tau_{1})},\end{eqnarray*}
where $t_{p}=\tau_{p}+\tau_{p-1}+\ldots+\tau_{1}+t_{0}$ is used.
Given the definition of the spectral waiting-time distribution\begin{equation}
\underline{w}(\omega)=\int_{0}^{\infty}d\tau w(\tau)e^{i\omega\tau},\label{eq:w-spectrum}\end{equation}
and the normalization of the waiting-time distribution, the $n$-fold
integral therefore results as

\begin{eqnarray}
 &  & \int_{t_{0}}^{\infty}dt_{n}\ldots\int_{t_{0}}^{t_{2}}dt_{1}w_{n}(t_{n},\ldots,t_{p},\ldots,t_{1})e^{2il\nu t_{p}}\nonumber \\
 &  & =\lambda_{2}\lambda_{1}^{n-1}e^{2il\nu t_{0}}\left[\underline{w}(2l\nu)\right]^{p}.\label{eq:n-fold-integral3}\end{eqnarray}

When Eq. (\ref{eq:n-fold-integral3}) is inserted into the first term
(\ref{eq:a2-mom-term1}), this term writes \begin{eqnarray*}
 &  & (-i\eta_{2})^{2l}\langle(\cos\theta)^{2l}\rangle_{2}e^{2il\nu t_{0}}\frac{\lambda_{2}}{\lambda_{1}}\sum_{n=1}^{\infty}\left[\lambda_{1}\underline{w}(2l\nu)\right]^{n}\\
 &  & =(-i\eta_{2})^{2l}\langle(\cos\theta)^{2l}\rangle_{2}e^{2il\nu t_{0}}\frac{\lambda_{2}\underline{w}(2l\nu)}{1-\lambda_{1}\underline{w}(2l\nu)}.\end{eqnarray*}
The sum of the other $n-1$ terms {[}Eq. (\ref{eq:a2-mom-otherterms})]
becomes\begin{eqnarray*}
 &  & (-i\eta_{1})^{2l}\langle(\cos\theta)^{2l}\rangle_{1}e^{2il\nu t_{0}}\sum_{n=1}^{\infty}\lambda_{2}\lambda_{1}^{n-1}\sum_{p=1}^{n-1}\left[\underline{w}(2l\nu)\right]^{p}\\
 &  & =(-i\eta_{1})^{2l}\langle(\cos\theta)^{2l}\rangle_{1}e^{2il\nu t_{0}}\frac{\lambda_{2}}{\lambda_{1}}\sum_{n=1}^{\infty}\lambda_{1}^{n}\frac{\underline{w}(2l\nu)-[\underline{w}(2l\nu)]^{n}}{1-\underline{w}(2l\nu)}\\
 &  & =(-i\eta_{1})^{2l}\langle(\cos\theta)^{2l}\rangle_{1}e^{2il\nu t_{0}}\frac{\lambda_{1}\underline{w}(2l\nu)}{1-\lambda_{1}\underline{w}(2l\nu)}.\end{eqnarray*}
Thus, the moment (\ref{eq:a2-mom}) of the density of recoil shifts
is \begin{eqnarray}
\langle\alpha^{2l}\rangle_{p} & = & (-1)^{l}\left(\eta_{2}^{2l}\langle(\cos\theta)^{2l}\rangle_{2}+\frac{\lambda_{1}}{\lambda_{2}}\eta_{1}^{2l}\langle(\cos\theta)^{2l}\rangle_{1}\right)\nonumber \\
 &  & \times\frac{\lambda_{2}\underline{w}(2l\nu)}{1-\lambda_{1}\underline{w}(2l\nu)}e^{2il\nu t_{0}}.\label{eq:B20}\end{eqnarray}

\subsection{Moment $\langle\alpha^{\ast2}\alpha^{2}\rangle_{p}$}

From Eq. (\ref{eq:P-moments2}) the moment $\langle\alpha^{\ast2}\alpha^{2}\rangle_{p}$
has the form\begin{eqnarray*}
 &  & \langle\alpha^{\ast2}\alpha^{2}\rangle_{p}=\sum_{n=1}^{\infty}\int_{t_{0}}^{\infty}dt_{n}\ldots\int_{t_{0}}^{t_{2}}dt_{1}w_{n}(t_{n},\ldots,t_{1})\\
 &  & \times\left\{ \partial_{\alpha}^{2}\partial_{\alpha^{\ast}}^{2}\left[\underline{p}_{2}(\alpha,t_{n})\underline{p}_{1}(\alpha,t_{n-1})\ldots\underline{p}_{1}(\alpha,t_{1})\right]\right\} _{\alpha=0}.\end{eqnarray*}
The non-vanishing derivatives acting on a single factor $\underline{p}_{a}(\alpha,t)$
are those of second and fourth order, i.e. the derivatives $\partial_{\alpha^{\ast}}^{2}\underline{p}_{a}$,
$\partial_{\alpha}^{2}\underline{p}_{a}$, $\partial_{\alpha}\partial_{\alpha^{\ast}}\underline{p}_{a}$,
and $\partial_{\alpha}^{2}\partial_{\alpha^{\ast}}^{2}\underline{p}_{a}$.
Therefore, the moment can be written as\begin{widetext}\begin{eqnarray}
\langle\alpha^{\ast2}\alpha^{2}\rangle_{p} & = & \sum_{n=1}^{\infty}\int_{t_{0}}^{\infty}dt_{n}\ldots\int_{t_{0}}^{t_{2}}dt_{1}w_{n}(t_{n},\ldots,t_{1})\left\{ \sum_{p=1}^{n}\partial_{\alpha}^{2}\partial_{\alpha^{\ast}}^{2}\underline{p}_{a_{p}}(\alpha,t_{p})\right.\nonumber \\
 &  & \left.+\sum_{p=1}^{n}\sum_{q\neq p}\left\{ \left[\partial_{\alpha^{\ast}}\partial_{\alpha}\underline{p}_{a_{p}}(\alpha,t_{p})\right]\left[\partial_{\alpha^{\ast}}\partial_{\alpha}\underline{p}_{a_{q}}(\alpha,t_{q})\right]+\left[\partial_{\alpha}^{2}\underline{p}_{a_{p}}(\alpha,t_{p})\right]\left[\partial_{\alpha^{\ast}}^{2}\underline{p}_{a_{q}}(\alpha,t_{q})\right]\right\} \right\} _{\alpha=0},\label{eq:a*2a2}\end{eqnarray}
\end{widetext}where $a_{p}=1$ for $p=1,\ldots,n-1$ and $a_{p}=2$
for $p=n$. The fourth-order derivatives that corresponds to the first
term in the braces, produce terms such as $\eta_{a}^{4}\langle(\cos\theta)^{4}\rangle_{a}$.
They occur once with $a=2$ and $n-1$ times with $a=1$, to yield,
after time integrations, the contribution\[
\eta_{2}^{4}\langle(\cos\theta)^{4}\rangle_{2}+\frac{\lambda_{1}}{\lambda_{2}}\eta_{1}^{4}\langle(\cos\theta)^{4}\rangle_{1}.\]
The next term in the braces of Eq. (\ref{eq:a*2a2}) results as the
time-independent double sum\begin{eqnarray*}
 &  & \sum_{p=1}^{n}\sum_{q\neq p}\eta_{a_{p}}^{2}\eta_{a_{q}}^{2}\langle(\cos\theta)^{2}\rangle_{a_{p}}\langle(\cos\theta)^{2}\rangle_{a_{q}}\\
 &  & =2\eta_{2}^{2}\eta_{1}^{2}(n-1)\langle(\cos\theta)^{2}\rangle_{2}\langle(\cos\theta)^{2}\rangle_{1}\\
 &  & \quad+\eta_{1}^{4}(n-1)(n-2)\langle(\cos\theta)^{2}\rangle_{1}^{2},\end{eqnarray*}
which, after time integrations in Eq. (\ref{eq:a*2a2}), adds to the
moment the contribution\begin{eqnarray*}
 &  & 2\eta_{2}^{2}\eta_{1}^{2}\left(\langle\hat{n}_{{\rm ph}}\rangle-1\right)\langle(\cos\theta)^{2}\rangle_{2}\langle(\cos\theta)^{2}\rangle_{1}\\
 &  & +\left[\eta_{1}^{4}\left(\langle\hat{n}_{{\rm ph}}\rangle-1\right)\left(\langle\hat{n}_{{\rm ph}}\rangle-2\right)+\eta_{1}^{4}\langle[\Delta\hat{n}_{{\rm ph}}]^{2}\rangle\right]\langle(\cos\theta)^{2}\rangle_{1}^{2}\\
 &  & =2\eta_{1}^{2}\frac{\lambda_{1}}{\lambda_{2}}\langle(\cos\theta)^{2}\rangle_{1}\left(\eta_{2}^{2}\langle(\cos\theta)^{2}\rangle_{2}+\frac{\lambda_{1}}{\lambda_{2}}\eta_{1}^{2}\langle(\cos\theta)^{2}\rangle_{1}\right).\end{eqnarray*}

The last term in the curved bracket of Eq. (\ref{eq:a*2a2}) produces
the double sum\begin{eqnarray*}
 &  & \sum_{p=1}^{n}\sum_{q\neq p}\eta_{a_{p}}^{2}\eta_{a_{q}}^{2}\langle(\cos\theta)^{2}\rangle_{a_{p}}\langle(\cos\theta)^{2}\rangle_{a_{q}}e^{2i\nu(t_{q}-t_{p})}\\
 &  & =2\Re\sum_{p=1}^{n}\sum_{q=1}^{p-1}\eta_{a_{p}}^{2}\eta_{1}^{2}\langle(\cos\theta)^{2}\rangle_{a_{p}}\langle(\cos\theta)^{2}\rangle_{1}e^{2i\nu(t_{p}-t_{q})},\end{eqnarray*}
which generates the contribution\begin{eqnarray}
 &  & 2\Re\sum_{n=1}^{\infty}\sum_{p=1}^{n}\sum_{q=1}^{p-1}\eta_{a_{p}}^{2}\eta_{1}^{2}\langle(\cos\theta)^{2}\rangle_{a_{p}}\langle(\cos\theta)^{2}\rangle_{1}\label{eq:last-contr}\\
 &  & \qquad\times\int_{t_{0}}^{\infty}dt_{n}\ldots\int_{t_{0}}^{t_{2}}dt_{1}w_{n}(t_{n},\ldots,t_{1})e^{2i\nu(t_{p}-t_{q})}.\nonumber \end{eqnarray}
The time integrals, \[
\int_{t_{0}}^{\infty}dt_{n}\ldots\int_{t_{0}}^{t_{2}}dt_{1}w_{n}(t_{n},\ldots,t_{1})e^{2i\nu(t_{p}-t_{q})},\]
with $p>q$, can be transformed as above into\begin{eqnarray*}
 &  & \lambda_{2}\lambda_{1}^{n-1}\int_{0}^{\infty}d\tau_{n}\ldots\int_{0}^{\infty}d\tau_{1}w(\tau_{n})\ldots w(\tau_{1})e^{2i\nu(\tau_{p}+...+\tau_{q+1})}\\
 &  & =\lambda_{2}\lambda_{1}^{n-1}[\underline{w}(2\nu)]^{p-q}.\end{eqnarray*}
Thus, the contribution (\ref{eq:last-contr}) becomes\begin{eqnarray*}
2\Re\sum_{n=1}^{\infty}\sum_{p=1}^{n}\sum_{q=1}^{p-1}\eta_{a_{p}}^{2}\eta_{1}^{2}\langle(\cos\theta)^{2}\rangle_{a_{p}}\langle(\cos\theta)^{2}\rangle_{1}\\
\times\lambda_{2}\lambda_{1}^{n-1}[\underline{w}(2\nu)]^{p-q}.\end{eqnarray*}
Using the substitution $q\to q^{\prime}=p-q$ and performing the sum
over $q^{\prime}$, we have\begin{eqnarray*}
 &  & 2\Re\sum_{n=1}^{\infty}\sum_{p=1}^{n}\eta_{a_{p}}^{2}\eta_{1}^{2}\langle(\cos\theta)^{2}\rangle_{a_{p}}\langle(\cos\theta)^{2}\rangle_{1}\lambda_{2}\lambda_{1}^{n-1}\\
 &  & \qquad\times\frac{\underline{w}(2\nu)}{1-\underline{w}(2\nu)}\left\{ 1-[\underline{w}(2\nu)]^{p-1}\right\} .\end{eqnarray*}
In the sum over $p$ we separate the term with $p=n$ from the terms
$p=1,\ldots,n-1$ and obtain\begin{eqnarray*}
 &  & 2\Re\sum_{n=1}^{\infty}\eta_{1}^{2}\langle(\cos\theta)^{2}\rangle_{1}\lambda_{2}\lambda_{1}^{n-1}\frac{\underline{w}(2\nu)}{1-\underline{w}(2\nu)}\\
 &  & \quad\times\Biggl\{\eta_{2}^{2}\langle(\cos\theta)^{2}\rangle_{2}\left\{ 1-\left[\underline{w}(2\nu)\right]^{n-1}\right\} \\
 &  & \qquad+\eta_{1}^{2}\langle(\cos\theta)^{2}\rangle_{1}\Biggl\{(n-1)-\frac{1-[\underline{w}(2\nu)]^{n-1}}{1-\underline{w}(2\nu)}\Biggr\}\Biggr\}.\end{eqnarray*}
The sum over $n$ produces terms like\[
\sum_{n=1}^{\infty}\lambda_{2}(\lambda_{1}x)^{n-1}=\frac{\lambda_{2}}{1-\lambda_{1}x},\]
where $x$ may be $1$ or $\underline{w}(2\nu)$. Thus, after performing
this sum, we arrive at \begin{widetext}\begin{eqnarray}
 &  & 2\Re\eta_{1}^{2}\langle(\cos\theta)^{2}\rangle_{1}\frac{\underline{w}(2\nu)}{1-\underline{w}(2\nu)}\nonumber \\
 &  & \times\left\{ \eta_{2}^{2}\langle(\cos\theta)^{2}\rangle_{2}\left[1-\frac{\lambda_{2}}{1-\lambda_{1}\underline{w}(2\nu)}\right]+\eta_{1}^{2}\langle(\cos\theta)^{2}\rangle_{1}\left\{ (\langle\hat{n}_{ph}\rangle-1)-\frac{1}{1-\underline{w}(2\nu)}\left[1-\frac{\lambda_{2}}{1-\lambda_{1}\underline{w}(2\nu)}\right]\right\} \right\} .\label{eq:contr-long}\end{eqnarray}
\end{widetext}The brackets can be simplified to\[
\left[1-\frac{\lambda_{2}}{1-\lambda_{1}\underline{w}(2\nu)}\right]=\frac{\lambda_{1}[1-\underline{w}(2\nu)]}{1-\lambda_{1}\underline{w}(2\nu)},\]
and $\langle\hat{n}_{{\rm ph}}\rangle=1/\lambda_{2}$, see Eq. (\ref{eq:n-avg}),
so that we obtain for the contribution (\ref{eq:contr-long})\begin{eqnarray*}
 &  & 2\eta_{1}^{2}\frac{\lambda_{1}}{\lambda_{2}}\langle(\cos\theta)^{2}\rangle_{1}\left(\eta_{2}^{2}\langle(\cos\theta)^{2}\rangle_{2}+\frac{\lambda_{1}}{\lambda_{2}}\eta_{1}^{2}\langle(\cos\theta)^{2}\rangle_{1}\right)\\
 &  & \times\Re\left[\frac{\lambda_{2}\underline{w}(2\nu)}{1-\lambda_{1}\underline{w}(2\nu)}\right].\end{eqnarray*}
The complete results is therefore:\begin{eqnarray}
\langle\alpha^{\ast2}\alpha^{2}\rangle_{p} & = & \left[\eta_{2}^{4}\langle(\cos\theta)^{4}\rangle_{2}+\frac{\lambda_{1}}{\lambda_{2}}\eta_{1}^{4}\langle(\cos\theta)^{4}\rangle_{1}\right]\nonumber \\
 &  & +2\eta_{1}^{2}\frac{\lambda_{1}}{\lambda_{2}}\langle(\cos\theta)^{2}\rangle_{1}\nonumber \\
 &  & \times\left(\eta_{2}^{2}\langle(\cos\theta)^{2}\rangle_{2}+\frac{\lambda_{1}}{\lambda_{2}}\eta_{1}^{2}\langle(\cos\theta)^{2}\rangle_{1}\right)\nonumber \\
 &  & \times\left[1+\Re\left[\frac{\lambda_{2}\underline{w}(2\nu)}{1-\lambda_{1}\underline{w}(2\nu)}\right]\right].\label{eq:a4-moment}\end{eqnarray}

\section{Spectral waiting-time distribution for resonant pumping \label{sec:waiting-time-distribution}}

The waiting-time distribution $w(t)$ is defined by\begin{equation}
w(t)=(\gamma_{1}+\gamma_{2})\left|\langle3|\hat{U}_{{\rm eff}}(t)|1\rangle\right|^{2},\end{equation}
{[}Eq. (\ref{eq:w-def})]. The time evolution $\hat{U}_{{\rm eff}}(t)$
is governed by the non-Hermitean Hamiltonian \begin{eqnarray}
\hat{H}_{{\rm eff}} & = & \frac{\hbar}{2}\left(\kappa\hat{\sigma}_{1,+}+\kappa^{\ast}\hat{\sigma}_{1,-}\right)+\hbar\Delta|1\rangle\langle1|\nonumber \\
 &  & -\frac{i\hbar}{2}\sum_{a=1,2}\gamma_{a}\hat{\sigma}_{a,+}\hat{\sigma}_{a,-},\end{eqnarray}
{[}Eq. (9)]. The transition amplitude $|1\rangle\to|3\rangle$ is
the projection of the solution of the equation of motion\begin{equation}
i\hbar\frac{\partial|\psi(t)\rangle}{\partial t}=\hat{H}_{{\rm eff}}|\psi(t)\rangle,\label{eq:dummy-schroedinger}\end{equation}
on state $|3\rangle$ with the initial condition $|\psi(t=0)\rangle=|1\rangle$.
Eq. (\ref{eq:dummy-schroedinger}) is decomposed in the set of coupled
equations for the components $\psi_{i}(t)=\langle i|\psi(t)\rangle$,
\begin{eqnarray}
\dot{\psi}_{1}(t) & = & -\frac{i\kappa^{\ast}}{2}\psi_{3}(t)-i\Delta\psi_{1}(t),\label{eq:coupled-eq1}\\
\dot{\psi}_{2}(t) & = & 0,\label{eq:coupled-eq2}\\
\dot{\psi}_{3}(t) & = & -\frac{i\kappa}{2}\psi_{1}(t)-\gamma\psi_{3}(t).\label{eq:coupled-eq3}\end{eqnarray}
where $\gamma=(\gamma_{1}+\gamma_{2})/2$. The second derivative $\ddot{\psi}_{3}(t)$
therefore results as\[
\ddot{\psi}_{3}(t)=-\frac{i\kappa}{2}\dot{\psi}_{1}(t)-\gamma\dot{\psi}_{3}(t),\]
which, with the use of Eq. (\ref{eq:coupled-eq1}), writes in the
resonant case ($\Delta=0$)

\begin{equation}
\left[\partial_{t}^{2}+\gamma\partial_{t}+\left|\frac{\kappa}{2}\right|^{2}\right]\psi_{3}(t)=0.\label{eq:combined}\end{equation}
 With the Ansatz \[
\psi_{3}(t)=\phi_{3}(t)\exp\left[-\frac{\gamma t}{2}\right],\]
the differential equation for $\phi_{3}(t)$ is obtained: \begin{equation}
\ddot{\phi}_{3}(t)=\left[\left(\frac{\gamma}{2}\right)^{2}-\left|\frac{\kappa}{2}\right|^{2}\right]\phi_{3}(t).\label{eq:ansatz}\end{equation}
The general solution is \begin{equation}
\phi_{3}(t)=ae^{\lambda t}+be^{-\lambda t},\label{eq:solution-general}\end{equation}
 with \begin{equation}
\lambda=\frac{\gamma}{2}\sqrt{1-S},\end{equation}
 where $S=|\kappa|^{2}/\gamma^{2}$ is the laser-saturation parameter. 

Thus, the time-evolved amplitude of state $|3\rangle$ reads\[
\psi_{3}(t)=\left(ae^{\lambda t}+be^{-\lambda t}\right)e^{-\gamma t/2}.\]
With the initial conditions $\psi_{3}(0)=0$ and $\psi_{1}(0)=1$,
one obtains, via Eq. (\ref{eq:coupled-eq3}), $\dot{\psi}_{3}(0)=-i\kappa/2$,
so that one finds the constants of integration, $a=-b=-i\kappa/(4\lambda)$.
Including these values, the above probability amplitude is\begin{equation}
\psi_{3}(t)=\frac{\kappa}{i\gamma\sqrt{1-S}}\sinh\left(\frac{\gamma t}{2}\sqrt{1-S}\right)e^{-\gamma t/2}.\label{eq:special-solution}\end{equation}
 The waiting-time distribution $w(t)=2\gamma|\psi_{3}(t)|^{2}$ is
then given by \begin{equation}
w(t)=\frac{2\gamma S}{\left|1-S\right|}\left|\sinh\left(\frac{\gamma t}{2}\sqrt{1-S}\right)\right|^{2}e^{-\gamma t}.\label{eq:P31-solution}\end{equation}
 Note that this distribution is normalized to unity: \[
\int_{0}^{\infty}dtw(t)=1.\]

The spectral waiting-time distribution results from the Fourier transform
of Eq. (\ref{eq:P31-solution}),\[
\underline{w}(\omega)=\int_{0}^{\infty}dtw(t)e^{i\omega t}.\]
It is obtained by straightforward calculation as\begin{eqnarray*}
\underline{w}(\omega) & = & \frac{\gamma^{3}S}{(\gamma-i\omega)\left[\gamma^{2}(S-1)+(\gamma-i\omega)^{2}\right]}\\
 &  & \times\left\{ \begin{array}{ll}
1, & S\leq1,\\
(-1), & S>1.\end{array}\right.\end{eqnarray*}
Thus, the required modulus and phase at twice the trap frequency become\begin{eqnarray}
\underline{w} & = & \frac{S\underline{w}^{({\rm sat})}}{\sqrt{(S-1)^{2}+2(S-1)(1-\tilde{\nu}^{2})+(1+\tilde{\nu}^{2})^{2}}},\label{eq:w-reson}\\
\tan\phi_{\underline{w}} & = & \frac{S+2-\tilde{\nu}^{2}}{S-3\tilde{\nu}^{2}}\tan\phi_{\underline{w}}^{({\rm sat})}\label{eq:phiw-reson}\end{eqnarray}
where the saturated values are\begin{eqnarray*}
\underline{w}^{({\rm sat})}=\lim_{s\to\infty}\underline{w} & = & 1/\sqrt{1+\tilde{\nu}^{2}},\\
\phi_{\underline{w}}^{({\rm sat})}=\lim_{s\to\infty}\phi_{\underline{w}} & = & \arctan\tilde{\nu},\end{eqnarray*}
with $\tilde{\nu}=2\nu/\gamma$.

\end{document}